\documentclass[a4paper,11pt]{article}
\pdfoutput=1
 \usepackage{jcappub}
\usepackage{enumerate}
\usepackage{epsfig} 
\usepackage{wasysym} 
\usepackage{mathrsfs} 
\usepackage{graphicx} 
\usepackage{caption}
\usepackage{subcaption}
\usepackage{amsfonts} 
\usepackage{amsbsy} 
\usepackage{amscd} 
\usepackage{pstricks} 
\usepackage{multirow} 

\newcommand{\lag}{\mathcal{L}}

\newcommand{\deriv}{\mathcal{D}}
\newcommand{\sm}{SU(3)_C\otimes SU(2)_L\otimes U(1)_Y}
\newcommand{\be}{\begin{equation}}
\newcommand{\ee}{\end{equation}}
\newcommand{\bea}{\begin{eqnarray}}
\newcommand{\eea}{\end{eqnarray}}

\title{Higgs vacuum stability and inflationary
  dynamics after BICEP2 and PLANCK dust polarisation data}

\author[a]{Kaushik Bhattacharya,}
\author[b]{Joydeep Chakrabortty,}
\author[c]{Suratna Das,} 
\author[d]{Tanmoy Mondal}
\affiliation[a,b,c]{Department of Physics, Indian Institute of Technology, Kanpur-208016, India}
\affiliation[d]{Theoretical Physics Division, Physical Research Laboratory, Ahmedabad-380009, India}
\affiliation[d]{Department of Physics, Indian Institute of Technology, Gandhinagar, Ahmedabad, India.}
\emailAdd{kaushikb@iitk.ac.in}
\emailAdd{joydeep@iitk.ac.in} 
\emailAdd{suratna@iitk.ac.in}
\emailAdd{tanmoym@prl.res.in}

\abstract
{
If the recent detection of $B-$mode polarization of the Cosmic Microwave
Background by BICEP2 observations, withstand the test of time after the 
release of recent PLANCK dust polarisation data, then it would surprisingly put the
inflationary scale near Grand Unification scale if one considers
single-field inflationary models. On the other hand, Large Hadron
Collider has observed the elusive Higgs particle whose presently
observed mass can lead to electroweak vacuum instability at high scale
$(\sim{\mathcal O}(10^{10})$ GeV). In this article, we seek for a simple
particle physics model which can simultaneously keep the vacuum of the
theory stable and yield high-scale inflation successfully. To serve
our purpose, we extend the Standard Model of particle physics with a
$U(1)_{B-L}$ gauged symmetry which spontaneously breaks down just
above the inflationary scale. Such a scenario provides a constrained
parameter space where both the issues of vacuum stability and
high-scale inflation can be successfully accommodated. The threshold
effect on the Higgs quartic coupling due to the presence of the heavy
inflaton field plays an important role in keeping the electroweak
vacuum stable. Furthermore, this scenario is also capable of reheating
the universe at the end of inflation. Though the issues of Dark Matter
and Dark Energy, which dominate the late-time evolution of our
universe, cannot be addressed within this framework, this model
successfully describes the early universe dynamics according to the
Big Bang model.
}

\begin{document}

\maketitle
%==============================================================================
\section{Introduction} 

The recent observations, in the fields of both particle physics and
cosmology, are immensely exciting as they have led to new discoveries
in their respective fields. On the one hand, in particle physics the
discovery of the elusive Higgs boson at the Large Hadron Collider
(LHC) \cite{Aad:2012tfa,Chatrchyan:2012ufa} completes the picture of
the Standard Model (SM) of particle physics. On the other hand, in the
field of cosmology the recent observation of $B-$mode polarization of
the Cosmic Microwave Background Radiation (CMBR) by BICEP2
\cite{Ade:2014xna}, if confirmed, would reveal the picture of the
primordial universe through the effect of tensor perturbations, which
had defied detection so far. 

We know new discoveries often lead to new conundrums and these
observations are no exceptions to that. They have also set profound
challenges in the field of particle physics. The observed mass of the
Higgs boson ($\sim 122-127$ GeV) at the LHC \cite{Aad:2012tfa,
  Chatrchyan:2012ufa} may lead to an unstable electroweak vacuum in
higher energy scales. This is because the scalar quartic coupling
eventually becomes negative at very high scale leaving the scalar
potential unbounded from below. Hence it is of importance to address
the issue of stability of the SM vacuum \cite{Coleman:1977py_sm_old}
which has been analyzed in \cite{Sher:1988mj, Sher:1993mf_sm_old,
  Espinosa:1995se_sm_old, Schrempp:1996fb_sm_old, Casas:1996aq_sm_old,
  Isidori:2001bm_sm_old, Holthausen:2011aa_smRGE, Alekhin:2012py_vs,
  Degrassi:2012ry_vs, Masina:2012tz_vs, Tang:2013bz_sm,Herranen:2014cua, Spencer-Smith:2014woa, Gabrielli:2013hma, Branchina:2013jra, Branchina:2014rva, Branchina:2014usa}. It has been
noted, that it is indeed a difficult task to maintain the stability of
the SM vacuum till the Planck scale \cite{Degrassi:2012ry_vs,
  Masina:2012tz_vs, Kobakhidze:2014xda}, which indicates towards the
onset of new physics before the electroweak vacuum becomes unstable
\cite{Kobakhidze:2014xda, Casas:1994qy_sm_old,
  Anchordoqui:2012fq_sm++, Hung:1979dn_sm_old, Chakrabortty:2013zja,
  u1b-l_recent, Chakrabortty:2013mha}.

Detection of $B-$mode polarization by BICEP2
experiment \cite{Ade:2014xna}, if confirmed, would put inflationary paradigm
\cite{Guth:1980zm, Linde:1983gd} on stronger footing than ever as only
the primordial tensor modes produced during inflation are capable of
imprinting $B-$mode polarization on the CMBR.  In the simplest
inflationary picture, the primordial universe during inflation is
dominated by the potential energy of a slowly rolling scalar field,
called inflaton, whose particle physics origin is yet to be
known. Non-detection of isocurvature modes and primordial
non-Gaussianity by the recent PLANCK observation \cite{Ade:2013zuv}
favour this simplest single-field model of inflation. The aim of the
BICEP2 experiment was to measure the ratio of the amplitudes of the
primordial tensor and the scalar perturbations, known as
the tensor-to-scalar ratio denoted by $r$. PLANCK put an upper limit 
on tensor-to-scalar ratio as $r<0.11$, 
whereas BICEP2 claimed to have measured
$r$ at large angular scales as
\begin{eqnarray}
r=0.2^{+0.07}_{-0.05},
\end{eqnarray}
where $r=0$ is ruled out at $7\sigma$ confidence level. Such dispute in observations can be 
overcome if in future the running of the scalar spectral index $(dn_s/d\ln k)$ turns out be 
large and negative \cite{Ade:2014xna}. Recently after the release 
of PLANCK's polarised dust data \cite{Adam:2014bub} the BICEP2's detection 
of $B$-mode polarisation has been put under severe scrutiny. Despite such doubt,  a
high-enough detection of $r$ has profound importance in cosmology for
many reasons. First of all, such a high value of $r$, as has been claimed to 
have been observed by BICEP2, indicates a
large-field inflationary scenario such as chaotic inflation
\cite{Linde:1983gd} and favours quadratic as well as quartic potentials of
the inflaton field \cite{Okada:2014lxa, Kobayashi:2014jga}. Secondly, the observed value of $r$ by BICEP2 sets the scale of
inflation surprisingly close to the Grand Unification (GUT) scale
($\sim10^{16}$ GeV). Furthermore, as one requires the dynamics of a
scalar field (preferably a fundamental scalar field) for inflation and
the only known candidate in nature so far is the Higgs field, it is
most economic to let the Higgs play the role of inflaton. Such Higgs
inflationary scenario was first proposed in
Ref.~\cite{Bezrukov:2007ep} where the Higgs field strongly couples to
the Ricci scalar. After the announcement of BICEP2 results, it has
been argued that detection of such high $r$ has made it difficult for
the SM Higgs boson to be the inflaton during inflationary era as these
scenarios generally produce low tensor mode amplitude $(r\sim{\mathcal
  O}(10^{-2}))$ \cite{Cook:2014dga, Fairbairn:2014nxa}.  There are
counter arguments, based on the roles of the top and the Higgs mass
during the inflationary period, to rescue the Higgs inflation scenario
\cite{Hamada:2014iga, Masina:2014yga, Bezrukov:2014bra}. To revive
such possibilities few attempts have been made by incorporating
non-minimal coupling between the Higgs kinetic term and the Higgs
field \cite{Nakayama:2014koa}, or the Einstein tensor
\cite{Germani:2014hqa} or both \cite{Oda:2014rpa}, or by adding a
cosmological constant \cite{Feng:2014naa} or by introducing extra
singlet scalar dark matter \cite{Haba:2014zda}.  From the discussion
so far we can infer that:
\begin{itemize}
\item The electroweak vacuum of the SM is not stable all the way up to
  the Planck scale if one does not invoke new physics in between the
  electroweak scale and the Planck scale.
\item In a generic Higgs inflationary scenario the SM Higgs field, not
  being able to produce high enough $r$ according to BICEP2 observations,
   is not favoured as inflaton.
  Thus one might take an alternative path to identify some other scalar
  field as inflaton, while keeping in mind that the SM Higgs
  essentially plays no dynamical role during inflation in order to
  yield a single-field inflationary epoch favoured by PLANCK.
\end{itemize} 
If one assumes that no other new physics sets in till inflationary
scale, which now according to BICEP2 observation takes place at the
GUT scale, then one needs to extend the SM scenario to the
inflationary scale where introducing new physics one can account for
the inflaton. It may happen that this new physics input in the GUT
scale can make the Higgs vacuum stable up to the Planck
scale \footnote{It has recently been observed that high-scale
  inflation, indicated by BICEP2 observations, would destabilize the
  SM Higgs vacuum during inflation and one thus requires to invoke
  small coupling between the inflaton and the SM Higgs field in order
  to avoid such catastrophe \cite{Fairbairn:2014zia, Lebedev:2012sy}.}. In this
article we will deal with a minimal extension of the SM which can
explain the vacuum stability of the SM as well as the inflationary
dynamics in the light of PLANCK and BICEP2 results. Furthermore, it is well known
that in non-supersymmetric theories the quadratic divergence of the
Higgs mass remains an open problem and few attempts have been made to
deal with such difficulties \cite{Hamada:2012bp}. In this article we
would also try to address whether the Higgs mass can be kept light at
the inflationary scale so that reheating at the end of
inflation can take place through the channel of inflaton decay into SM
Higgs field. There are few other very important outstanding issues
with the present Standard Models of particle physics and cosmology,
such as the origin of Dark Energy and Dark Matter, which are beyond
the scope of the present article.

To serve our purpose, we extend the SM by an additional $U(1)_{B-L}$
gauge symmetry. Phenomenological aspects and the issues related to the
vacuum stability of this $U(1)_{B-L}$ extended SM have been
extensively analyzed in \cite{Marshak:1979fm_BL,
  Buchmuller:1991ce_U1BL, Emam:2007dy_Zprime, Khalil:2010iu_tevBL,
  Iso:2009nw_tevBL, Basso:2008iv_phenoBL, Chakrabortty:2013zja,
  u1b-l_recent}. Such extension of SM has also been discussed
previously in several cosmological contexts such as in inflationary
scenarios \cite{Okada:2011en, Okada:2013vxa}, to explain the origin of
dark matter \cite{Arcadi:2013qia, Sanchez-Vega:2014rka,
  Basak:2013cga, Khalil:2008kp, Okada:2010wd, Kanemura:2011vm,Basso:2012ti}, baryogenesis and leptogenesis \cite{Perez:2014qfa,
  Hook:2014mla, Basso:2012ti} and production of gravitational waves
\cite{Buchmuller:2013lra}. Here we consider the spontaneous breaking
of the $U(1)_{B-L}$ symmetry and the electro-weak symmetry to take
place at very different energy scales. While the former takes place
above the inflationary scale $(\sim 10^{16}\,{\rm GeV})$ and the real
part of the scalar of this symmetry group plays the role of inflaton,
the SM electroweak symmetry breaking takes place around 246 GeV. Also,
to our advantage, couplings between the scalar of the $U(1)_{B-L}$ and
the SM particles help reheat the universe at the end of inflation.

We have organized the rest of the article as follows. In
Section~\ref{mod} we explain our particle physics model in the
cosmological backdrop. In the Section~\ref{res} the constraints on
model parameters from vacuum stability and inflationary dynamics are
presented and the last section summarizes the main results obtained
in the present work.
%=============================================================================
\section{The model}
\label{mod}
In this section we discuss the particle physics model, the SM extended
by an Abelian $U(1)_{B-L}$ group, which we will pursue to unify the
concepts of inflation and electroweak vacuum stability.
%---------------------------------------------------------------------------- 
\subsection{Scalar sector}
The $U(1)_{B-L}$ gauged extended SM, where the full symmetry group is
depicted as
\begin{equation}
\sm\otimes U(1)_{B-L},
\end{equation}
contains three extra right handed neutrinos to cancel all the gauge as
well as gravitational anomalies, one extra gauge boson and one extra
heavy scalar field $(\Phi)$ along with the SM particles. Here the
complex scalar field $\Phi$, which is singlet under SM but carries a
non-zero $B-L$ charge, is required to break the $U(1)_{B-L}$ symmetry
just above the inflation scale, and after the symmetry breaking the
real part of $\Phi$ is identified as the inflaton in our scenario.
Once this field $\Phi$ acquires the vacuum expectation value (vev),
$U(1)_{B-L}$ symmetry is spontaneously broken around $10^{16}$ GeV, a
scale which is governed by the scale of inflation as suggested by BICEP2
measurement \cite{Kehagias:2014wza}. The existence of the SM Higgs
doublet ensures the breaking of electroweak symmetry at $\sim 246$
GeV.\footnote{We have defined the vev as $v/\sqrt{2}$.}  The
important point to note here is that in such a scenario symmetry
breaking of $U(1)_{B-L}$ and the SM electroweak vacuum take place at
two very different energy scales.

The part of the Lagrangian that contains the kinetic and potential
terms of the scalars present in this theory is expressed as:
\begin{eqnarray} \label{eqn:lagrangian}
\lag & = & (\deriv_\mu S)^{\dagger} (\deriv^\mu S) +
(\deriv_\mu \Phi)^{\dagger} (\deriv^\mu \Phi) \nonumber + m_s^2 (S^{\dagger} S) + m_\phi^2 (\Phi^{\dagger} \Phi) \nonumber \\ &  & - \lambda_1
(S^{\dagger} S)^2 - \lambda_2 (\Phi^{\dagger} \Phi)^2 -
\lambda_3 (S^{\dagger} S) (\Phi^{\dagger} \Phi),
\end{eqnarray} 
where $S$ represents the SM Higgs field,{\footnote{ As our work
    includes both particle physics and cosmology where the Higgs field
    and the Hubble parameter, both conventionally represented by $H$,
    have important roles to play, we choose the convention where the
    Hubble parameter is represented by $H$ and the SM Higgs field as
    $S$ to avoid confusion.} $m_s$ and $m_\phi$ are the mass
  parameters of the Higgs and the $U(1)_{B-L}$ scalar fields
  respectively. Here, $\lambda_{1}$ and $\lambda_2$ are the quartic
  couplings of $S$ and $\Phi$ respectively and $\lambda_3$ is the
  coupling between $S$ and $\Phi$.  In the unbroken $SU(2)_L \otimes
  U(1)_{Y} \otimes U(1)_{B-L}$ theory the covariant derivative is
  defined as:
\begin{eqnarray}
\deriv^\mu \equiv \partial^\mu + i g_2 \tau_j W^\mu_j
+ i g_1 Y B_1^\mu + i (g' Y + g_{B-L} Q_{B-L}) B_2^\mu,
\end{eqnarray}
where $B_2^\mu$ is the $B-L$ charged gauge bosons whose kinetic term
is given as
\begin{eqnarray} 
\lag_{K.E.}^{B-L} = \frac{1}{4} (\partial^\mu B_2^\nu -
\partial^\nu B_2^\mu) (\partial_\mu B_{2 \nu} - \partial_\nu B_{2
  \mu}).  
\end{eqnarray}
Just above the inflationary scale, spontaneous breaking of the
$U(1)_{B-L}$ symmetry yields $\Phi
=\frac{1}{\sqrt{2}}(v_\phi+\phi(t,{\mathbf x}))$ where $v_\phi
\equiv\sqrt{m_\phi^2/\lambda_2}$ is the vev acquired by $\Phi$ and we
have not written the phase part which yields the Goldstone mode. The
real part, $\phi(t,{\mathbf x})$, of $\Phi$, apart from the vev, can
be written as a background field $\phi_0(t)$ which plays the role of
inflaton and fluctuations $\delta\phi(t, \mathbf{x})$ which give rise
to the primordial perturbations during inflation. After the
spontaneous breaking of $B-L$ symmetry, the scalar potential
$V(S,\Phi)$ of this extended theory  can be written as :
\begin{eqnarray}
V(S,\Phi)=\lambda_1(S^\dagger S)^2-m_s^2(S^\dagger S)+\lambda_2\left(\Phi^\dagger\Phi-\frac12v_\phi^2\right)^2+\lambda_3(S^\dagger S)\left(\Phi^\dagger\Phi-\frac12v_\phi^2\right).
\label{u1-ssb}
\end{eqnarray}
We see that various possible terms are generated in the scalar
potential part of the Lagrangian, like $\lambda_3\,v_\phi^2\,
S^{\dagger} S$, $\lambda_3\,v_\phi\,S^{\dagger}S\,\phi_0$, $\lambda_3
\,S^{\dagger} S\,\phi_0 \phi_0$. The first term redefines the mass
parameter of the $S$ field, the second term opens up the possibility
of decay of inflaton into two SM Higgs fields during reheating. The
third term introduces scattering of the light Higgs and the inflaton
during the inflationary regime.  We will concentrate on the importance
of the second term later while discussing the decay of inflaton during
reheating.

After the electroweak symmetry is broken at $246\,{\rm GeV}$ the Higgs
field is redefined as $S= \left(0
~~~~\frac{1}{\sqrt{2}}(v_s+s)\right)^{T}$ and below this scale both
the scalar fields get mixed and the physical fields ($\phi_l$ and
$\phi_h$, where the subscripts $l$ and $h$ stands for `light' and
`heavy' respectively) are achieved by diagonalizing the scalar mass
matrix. The physical masses of these scalars are given as
\begin{eqnarray}
M_{\phi_l,\phi_h}^2 = \frac{1}{2}\left[ \lambda_1 v_s^2 +
  \lambda_2 v_\phi^2 \pm \sqrt{(\lambda_1 v_s^2 - \lambda_2 v_\phi^2)^2 +
    \lambda_3^2 v_s^2 v_\phi^2} \right],
\label{phy-scalar-mass}
\end{eqnarray}
where the mixing angle is 
\begin{eqnarray}
\tan (2\alpha) = \frac{\lambda_3 v_s v_\phi} 
{\lambda_1 v_s^2 -\lambda_2 v_\phi^2}.
\label{mixing-angle}
\end{eqnarray}
We have set no Abelian mixing at tree level, i.e., $g'=0$ at
electroweak scale which can be done without any loss of
generality. But this mixing will arise through the renormalisation
group evolutions \cite{delAguila:1995rb_2U1, Holdom:1985ag_2U1,
  delAguila:1988jz_2U1} and that has been taken into account in our
analysis.  Within this framework the gauge boson masses are
\begin{eqnarray}
M_Z^2= \frac{1}{4}(g_1^2 + g_2^2) v_h^2\,,\,\,\,\,\,
M_{Z_{B-L}}^2=4 g_{B-L}^2 v_\phi^2.
\end{eqnarray}

One of our aims in this article is to seek for stability of
electroweak vacuum all the way up to the Planck scale. We know the
structure of the scalar potential is determined by the quartic
couplings ($\lambda_i$) at large field values. Thus the $\lambda_i$s
determine whether the potential is bounded from below or not. To
achieve a stable vacuum the potential needs to be bounded from below
at all energy scales, and this is a sufficient condition. Now, as
stated before, the energy scales of the $U(1)_{B-L}$ gauged extended
SM and the low energy effective theory, which is just the SM, are well
separated in our case. Thus the `decoupling theorem'
\cite{Appelquist:1974tg} states that the effects due to the new heavy
particles of the extended theory would not affect the quartic coupling
of the SM Higgs at low scales. This implies that at lower scales
evolution of Higgs quartic coupling is governed by the SM particles
only. Then it might indicate that extending the SM symmetry group
would have no effect in stabilizing the electroweak vacuum at lower
scales as the extended theory and the low energy effective SM theory
are `decoupled'. But the advantage of extending the SM symmetry group
with such a high scale $U(1)_{B-L}$ gauge symmetry is that the
extended theory contains a heavy scalar, which plays the role of an
inflaton in our scenario. As has been proposed in
\cite{EliasMiro:2012ay, Lebedev:2012zw}, presence of a heavy scalar, besides the SM
particles, eventually leads to a threshold correction to the SM Higgs
quartic coupling and helps stabilize the electroweak vacuum as long as
the mass of the heavy scalar lies below the instability scale of
electroweak vacuum which is around $10^{10}$ GeV. This is the key feature
of our model and we would show that though one requires to break the
$U(1)_{B-L}$ symmetry at very high scale ($\sim10^{16}$ GeV) to have
successful inflation at GUT scale, the mass of the inflaton can lie
below the electroweak instability scale as the quartic coupling of the
inflaton has to be fine tuned to yield the correct amplitude of scalar
power spectrum, as we show below.

To show how the threshold correction, due to presence of a heavy
scalar, modifies the evolution of Higgs quartic coupling $\lambda_1$
at lower scale \cite{EliasMiro:2012ay, Lebedev:2012zw}, let us consider the scalar
potential after $U(1)_{B-L}$ symmetry breaking given in
Eqn.~(\ref{u1-ssb}). At lower energy scales, when the heavy scalar
$\Phi$ has reached its minima, its equation of motion yields
\begin{eqnarray}
\Phi^\dagger\Phi=\frac12 v_\phi^2-\frac{\lambda_3}{2\lambda_2}S^\dagger S.
\end{eqnarray}
Below the mass scale $m_\phi$ of the inflaton, one can thus integrate
out the heavy field $\Phi$ using the above equation of motion and the
potential given in Eqn.~(\ref{u1-ssb}) becomes
\begin{eqnarray}
V(S)|_{\rm eff}=\left(\lambda_1-\frac{\lambda_3^2}{4\lambda_2}\right)(S^\dagger S)^2-m_s^2(S^\dagger S).
\label{Veff}
\end{eqnarray}
After that the dynamics of this theory is effectively governed by
the SM particles where SM scalar potential is written as
\begin{eqnarray}
V(S)|_{\rm SM} \equiv \lambda_S(S^\dagger S)^2-m_s^2(S^\dagger S).
\end{eqnarray}
Here $\lambda_S$ is the SM Higgs quartic coupling related to the
electroweak symmetry breaking scale and the SM Higgs mass only. At
$m_{\phi}$ scale the impact of heavy inflaton field redefines the
Higgs quartic coupling as
$\lambda_S=(\lambda_1-\frac{\lambda_3^2}{4\lambda_2})$.  This is a
pure tree-level effect by which the heavy scalar of the extended
theory affects the stability bound of the low energy effective theory
even when these two theories are effectively decoupled. The Higgs
quartic coupling $\lambda_S$ of the low energy effective theory
receives a positive shift at the mass scale of the inflaton which thus
helps avoid the instability which might have occurred above $m_\phi$
scale.

%==============================================================================

\subsection{Inflation in the extended model}

In this extended model under consideration the real part of the
$U(1)_{B-L}$ breaking scalar field, i.e., $\phi_0$, plays the role of
inflaton. Such a scenario has previously been considered in
\cite{Okada:2013vxa}. Before going into the details and
particularities of our inflationary set up, we first briefly discuss
what we know from the simplest single-field inflationary model in the
light of recent BICEP2 as well as PLANCK observations. The amplitude
of the two-point correlation function or the power spectrum of
primordial scalar perturbations are measured through the two-point
correlation of the temperature fluctuations in the CMBR.  PLANCK has
measured this value as \cite{Ade:2013zuv}
\begin{eqnarray}
{\mathcal P}_{\mathcal R}\sim 2.215\times 10^{-9}.
\end{eqnarray}
The ratio of the tensor (${\mathcal
  P}_T$) and the scalar (${\mathcal P}_{\mathcal R}$) power spectrum is represented as
\begin{eqnarray}
r=\frac{{\mathcal P}_T}{{\mathcal P}_{\mathcal R}}\,,
\end{eqnarray}
where $r$ is conventionally called the tensor-to-scalar ratio.  This
ratio $r$ has recently been measured by the BICEP2 experiment to be
$0.20^{+0.07}_{-0.05}$ \cite{Ade:2014xna}. But after the release of
PLANCK's recent dust data \citep{Adam:2014bub} the observation of BICEP2 has
been put under serious scrutiny. Though for the time being, before PLANCK
and BICEP2 combine their observations, the
upper-limit on $r$ set by PLANCK \cite{Adam:2014bub} still survives, i.e.,
\begin{eqnarray}
r<0.11\quad (95\%\,{\rm CL}).
\end{eqnarray}
In single-field scenarios, the tensor power spectrum turns out to
be a sole function of the Hubble parameter $H$ during inflation as
\begin{eqnarray}
{\mathcal P}_T=\frac{2}{\pi^2}\frac{H^2}{M_{\rm Pl}^2},
\end{eqnarray}
where $M_{\rm Pl}\sim2.4\times10^{18}$ GeV is the reduced Planck
mass. As the Hubble parameter during inflation is related to the
inflaton potential $V_\phi$ in the following way
\begin{eqnarray}
H^2=\frac{V_\phi}{3M_{\rm Pl}^2},
\label{friedmann}
\end{eqnarray}
knowing ${\mathcal P}_{\mathcal R}$ and $r$ one can determine
both the Hubble parameter during inflation
\begin{eqnarray}
H\sim\sqrt{r}\times10^{-4}M_{\rm Pl},
\end{eqnarray}
and the scale of inflation $V_\phi^{1/4}$ 
\begin{eqnarray}
V_\phi^{\frac14}\sim\left(\frac{r}{0.01}\right)^{\frac14}\times10^{16}\,{\rm GeV}.
\end{eqnarray}
Furthermore, in a single-field model the scalar power spectrum turns
out to be
\begin{eqnarray}
{\mathcal P}_{\mathcal R}=\frac{H^2}{(2\pi)^2}\left(\frac{H^2}{\dot\phi_0^2}\right),
\label{power-scalar}
\end{eqnarray}
where the over-dot represents derivative with respect to cosmic time
$t$ and this yields the tensor-to-scalar ratio as
\begin{eqnarray}
r=\frac{8}{M_{\rm
    Pl}^2}\left(\frac{\dot\phi_0}{H}\right)^2=\frac{8}{M_{\rm
    Pl}^2}\left(\frac{d\phi_0}{dN}\right)^2,
\end{eqnarray}
where $N$ is the number of
$e$-foldings during inflation. This indicates that the excursion of
the inflaton field during inflation would be
\begin{eqnarray}
\frac{\Delta\phi_0}{M_{\rm Pl}}=\int_{N_{\rm end}}^{N_{\rm CMB}}dN\sqrt{\frac r8}\,,
\end{eqnarray}
where $N_{\rm end}$ and $N_{\rm CMB}$ are the number of $e$-foldings
at the end of inflation and when the largest observable mode in the
CMBR leave the horizon before inflation ends, respectively.  Assuming
that $r$ would not change much during inflation, and $\Delta N\approx
65$ to solve the issues with Big Bang scenario, we have
\begin{eqnarray}
\frac{\Delta\phi_0}{M_{\rm Pl}}=\sqrt{530\times r}.
\end{eqnarray}
Hence, for $r\geq \mathcal{O}(10^{-2})$  the field excursion during inflation would be
super-Planckian (large-field inflationary models), and for $r < \mathcal{O}(10^{-2})$ it
would be sub-Planckian (small-field inflationary models).

In the present model the inflaton potential can be written as
\cite{Okada:2013vxa}
\begin{eqnarray}
V(\phi_0)=\frac{1}{4}\lambda_2(\phi_0^2-v_\phi^2)^2
+ a\lambda_2\log\left(\frac{\phi_0}{v_\phi}\right)\phi_0^4,
\label{full-potential}
\end{eqnarray}
where we have 
\begin{eqnarray} \label{eqn:radiative-correction}
a \equiv
\frac{1}{16\pi^2\lambda_2}\left(20\lambda_2^2+2\lambda_3^2+2\lambda_2
\left(\sum_i(Y_i^{N_R})^2-24g^2_{B-L}\right)+96g^4_{B-L}-
\sum_i(Y_i^{N_R})^4\right).\;\;
\label{rad-a}
\end{eqnarray}
The above potential contains the radiative correction added to the
tree-level one. Here $Y_i^{N_R}$ stand for the right handed neutrino
Yukawa couplings. The value of `$a$' determines whether the
$U(1)_{B-L}$ symmetry is broken through the tree-level potential or
the radiatively generated logarithmic term. As the value of `$a$'
mostly depends on the value of $g_{B-L}$ and $Y_i^{N_R}$, it can
either be positive or negative depending upon the values of the
couplings at inflationary scale. At tree level one can then identify
the mass term of the inflaton as
\begin{eqnarray}\label{eqn:mphi0}
m_\phi=\sqrt{\lambda_2}v_{\phi}\,.
\label{m-v}
\end{eqnarray}
In large-field inflationary models one would naturally expect the
quartic term with radiative corrections to dominate over the mass term
in the inflaton potential and the form of the potential which would be
responsible for inflation will be
\begin{eqnarray}
V_\phi(\phi_0)\approx\frac{1}{4}\lambda_2\phi_0^4+ a\lambda_2\log\left(\frac{\phi_0}{v_\phi}\right)\phi_0^4.
\end{eqnarray}
The flatness of the potential is determined by the
slow-roll parameters defined as
\begin{equation}
\epsilon_V = \frac12 M_{\rm Pl}^2\left(\frac{V'_\phi}{V_\phi}\right)^2, \;\;\;
\eta_V = M_{\rm Pl}^2\left(\frac{V_\phi''}{V_\phi}\right),\;\;\;\xi_V^2=M_{\rm Pl}^4\left(\frac{V_\phi'V_{\phi}'''}{V_\phi^2}\right),
\end{equation}
where the prime denotes derivative with respect to the field
$\phi_0$. These slow-roll parameters remain small ($\epsilon_V,
\eta_V\ll1$) during inflation till $\epsilon_V$ becomes $\sim1$, which
marks the end of inflation.  In a single-field scenario the
tensor-to-scalar ratio $r$ and the scalar spectral index $n_s$ (which
is a measure of the tilt of the scalar power spectrum) are related to
the slow-roll parameters in the following way
\begin{eqnarray}
r\approx 16\epsilon_V,\quad\quad n_s\approx1-6\epsilon_V+2\eta_V,
\end{eqnarray}
which are measured in CMBR observations. PLANCK measures the scalar
spectral index as $n_s=0.9603\pm0.0073$ \cite{Ade:2013uln}. The
evolution of the scalar spectral index can also be determined in terms
of the slow-roll parameters as:
\begin{eqnarray}
\frac{dn_s}{d\ln k}\approx16\epsilon_V\eta_V-24\epsilon_V^2-2\xi_V^2.
\end{eqnarray}

If one assumes that the  quartic self-interacting term without the radiative correction in the inflaton potential
  drives inflation, the tensor-to-scalar 
ratio and the scalar spectral index turn out to be
\begin{eqnarray}
r&=&\frac{128M_{\rm Pl}^2}{\phi_0^2},\nonumber\\
n_s&=&1-\frac{24M_{\rm Pl}^2}{\phi_0^2}.
\end{eqnarray}
The number of $e$-foldings can be computed as
\begin{eqnarray}
N_k=\frac{1}{M_{\rm Pl}^2}\int_{\phi_{0_{\rm end}}}^{\phi_{0_k}}\frac{V_\phi d\phi_0}{V_\phi'},
\end{eqnarray}
where $\phi_{0_k}$ is the field value at the co-moving scale $k$ and
$\phi_{0_{\rm end}}$ is the field value at the end of inflation. This yields
\begin{eqnarray}
N_{k_*}\approx\frac{\phi_{0_{k_*}}^2}{8M_{\rm Pl}^2},
\label{deltaN}
\end{eqnarray}
where $k_*$ denotes the pivot scale and it has been considered that
$\phi_{0_{\rm end}}\ll\phi_{0_{k_*}}$. Hence, if the mode
corresponding to the pivot scale would have left the horizon around 65 $e$-foldings
before inflation ends, then the above expression helps to determine the
field value during that time, which turns out to be
$\phi_{0_{k_*}}\sim 23 M_{\rm Pl}$. The tensor-to-scalar ratio and the
scalar spectral index at scale $k$ can be expressed in terms of the $e$-foldings as
\begin{eqnarray}
r_k&=&\frac{16}{N_k},\nonumber\\
n_{s_k}&=&1-\frac{3}{N_k}.
\end{eqnarray}
The number of $e$-folds when the pivot scale crosses the horizon during
inflation can also be written as
\begin{eqnarray}
N_*\simeq65+2\ln\left(\frac{V_\phi(\phi_{0_*})^{1/4}}{10^{14}\,{\rm GeV}}\right)-\ln\left(\frac{T_f}{10^{10}\,{\rm GeV}}\right),
\label{pivot-efold}
\end{eqnarray}
where $T_f$ is the temperature at the end of inflation and  can be
considered as the reheating temperature $T_{\rm RH}$. If the pivot scale
set by PLANCK, i.e., $k_*=0.002$ Mpc$^{-1}$, crosses the horizon during
inflation when $N_*\sim65$ then it generates large tensor-to-scalar ratio
as $r_*\sim0.25$ which is also large enough even for BICEP2
observations. This corresponds to the field excursion during inflation
to be $\Delta\phi\sim12\,M_{\rm Pl}$. Hence, our aim would be to
generate lower values of $r$ while keeping the scenario consistent
with the observations of $n_s$ and ${\mathcal P}_{\mathcal R}$ by
PLANCK. It has been pointed out in \cite{NeferSenoguz:2008nn} that the
radiative corrections to the quartic potential play an important role
to lower the tensor-to-scalar ratio.  Hence, for our inflationary
scenario we consider the  inflaton potential including radiative correction for
inflation.  When inflation is driven by this quartic potential, we find
\begin{eqnarray}\label{full_potential}
V_\phi&=&\frac14\lambda_2\phi_0^4\left[1+4a\ln\left(\frac{\phi_0}{v_\phi}\right)\right],\nonumber\\
V_\phi'&=&\lambda_2\phi_0^3\left[1+a+4a\ln\left(\frac{\phi_0}{v_\phi}\right)\right],\nonumber\\
V_\phi''&=&3\lambda_2\phi_0^2\left[1+\frac73a+4a\ln\left(\frac{\phi_0}{v_\phi}\right)\right],\nonumber\\
V_\phi'''&=&6\lambda_2\phi_0\left[1+\frac{13}{3}a+4a\ln\left(\frac{\phi_0}{v_\phi}\right)\right].
\end{eqnarray}
These give the slow-roll parameters as
\begin{eqnarray}
\epsilon_V &=& \frac{8M_{\rm Pl}^2}{\phi_0^2}\Big[\frac{u^2}{(u-1)^2}\Big],~~
\eta_V  =  \frac{12M^2_{\rm Pl}}{\phi_0^2} \Big[\frac{u+4/3}{u-1}\Big],~~
\xi_V^2 = \frac{96M_{\rm Pl}^4}{\phi_0^4} \Big[ \frac{u(u+10/3)}{(u-1)^2}\Big].
\label{slow-rolls}
\end{eqnarray}
where we have defined $u=(1+a+4a\ln{\phi_0/v_\phi})/a$. Hence, the
tensor-to-scalar ratio, the scalar spectral index and the running of
the scalar spectral index can be written as:
\begin{eqnarray}
r&=&\frac{128M_{\rm Pl}^2}{\phi_0^2}\frac{u^2}{(u-1)^2},\nonumber\\
n_s&=&1-\frac{8M_{\rm Pl}^2}{\phi_0^2}\frac{3u^2-u+4}{(u-1)^2},\nonumber\\
\frac{dn_s}{d\ln k}&=&-\frac{64M_{\rm Pl}^4}{\phi_0^4}\left[\frac{u(3u^3-4u^2+15u+10)}{(u-1)^4}\right],
\end{eqnarray}
respectively.
If we consider the radiative correction to the scalar potential as given in Eqn.~(\ref{full_potential}) is negligible which implies $a\rightarrow0$, i.e., $u\rightarrow\infty$, the standard expressions for $r$ and $n_s$ for quartic coupling can be recovered.
Furthermore, the scalar power spectrum given in
Eqn.~(\ref{power-scalar}), can be represented as
\begin{eqnarray}
{\mathcal P}_{\mathcal
  R}=\frac{1}{24\pi^2\epsilon_V}\left(\frac{V_\phi}{M_{\rm Pl}^4}\right)=\frac{1}{12\pi^2M_{ \rm Pl}^6}\left(\frac{V_\phi^3}{V_\phi^{'2}}\right),
  \label{scalar-power1}
\end{eqnarray}
where  we have used the Hubble slow-roll
parameter $\epsilon=\frac{1}{2M_{\rm Pl}^2}\frac{\dot\phi_0^2}{H^2}$
and the Friedman equation during inflation given in
Eqn.~(\ref{friedmann}) with $\epsilon\approx\epsilon_V$.  The power
spectrum for the inflaton potential including radiative correction turns out to be
\begin{eqnarray}
 {\mathcal P}_{\mathcal R}
=\frac{\lambda_2}{768\pi^2}\left(\frac{\phi_0}{M_{\rm Pl}}\right)^6\frac{a(u-1)^3}{u^2}.
\end{eqnarray}
Now, this scenario can be realised in two cases.  In the limit $u\gg1$,
one can have $|a|\ll1$, then the radiative corrections become
negligible. In such a case the standard results for $\phi^4$ potential
should be retrieved. The other branch known as
hilltop solution is important when $u\approx1$ leading to
$a\sim-\left(4\ln(\phi_0/v_\phi)\right)^{-1}$.

We also require to determine the reheat temperature in order to
compute the number of e-foldings which corresponds to the pivot scale
as given in Eq.~(\ref{pivot-efold}). We notice that, apart from the
self-interaction term, the inflaton field is also coupled to the SM
Higgs field via the mixing term $\lambda_3$ which allows it to decay
into a pair of SM Higgs during inflation. The decay rate of such an
interaction is given as \cite{Okada:2013vxa}:
\begin{eqnarray}
\Gamma_S(\phi_0\rightarrow SS)=
\frac{\lambda_3^2v_\phi^2}{32\pi m_\phi}.
\label{gss}
\end{eqnarray}
This decay of inflaton field into SM Higgs would make inflaton
unstable for larger values of $\lambda_3$. Thus one requires to
restrain the decay width of the inflaton during inflation. This
requirement can be met if one demands that $\Gamma_S < m_\phi$ which
yields
\begin{eqnarray}
\lambda_3< \sqrt{32\pi\lambda_2}.
\end{eqnarray}
From Eqn.~(\ref{gss}) we can also roughly estimate the order of
reheating temperature $T_{\rm RH}$ if the reheating phase is dominated
by the Higgs decay. If during the reheating phase the inflaton and its
decay products are just in equilibrium then $\Gamma_S \sim H$ where 
$H$ is the Hubble parameter during the radiation dominated reheating
phase. This condition yields
\begin{eqnarray}
\frac{\lambda_3^2v_\phi^2}{32\pi m_\phi}=\sqrt{\frac{\pi^2}{90}g_*}\,\, 
\frac{T_{\rm RH}^2}{M_{\rm Pl}}\,,
\label{trhb}
\end{eqnarray}
where $g_* \sim 100$.

Now, let us determine the parameters for a large-field inflationary
scenario and take $\phi_{0_{k_*}}\sim 23 M_{\rm Pl}$. Putting the
central value of scalar spectral index as $n_s=0.9603$ we find two
solutions ($u_*$) for $u$ at the pivot scale: $-0.333$ and
$-11.001$. The first solution indicates a hilltop branch inflation
whereas the second one gives rise to a $\phi^4-$branch
inflation. Let us now analyse the parameters for these two scenarios :
\begin{itemize}
\item {\bf Hilltop inflation :} If one sets the vev that breaks $U(1)_{B-L}$, i.e., the scale of inflation inflation as
  $10^{16}$ GeV, then for $u_*=-0.333$ one finds $a_*\sim-0.028$. This
  indicates the field value at the end of inflation would be $\phi_{0_{\rm end}}\sim0.71\,M_{\rm
    Pl}$ when  $\epsilon_V\approx1$.  This value of $u_*$ yields the tensor-to-scalar ratio as
  $r_*=0.015$ and the inflaton quartic coupling, from the observation
  of the scalar power amplitude by PLANCK, as
  $\lambda_2\sim1.89\times10^{-13}$.  This yields the tree-level
  mass of the inflaton as $m_\phi\sim4.3\times10^9$ GeV. The evolution
  of the spectral index in such a scenario would be $\frac{dn_s}{d\ln
    k}|_{k_*}\sim 1.07\times10^{-4}$. In this scenario the
  inflaton-Higgs coupling can be of the order of $\sim10^{-6}$, which
  yields the reheating temperature as $T_{\rm RH}\sim1.29\times10^{13}$
  GeV. This reheating temperature and the energy-scale of inflation yield
  the $e$-folding at which pivot scale would have exited the horizon as
  $N_*\sim67$. Fig.~(\ref{hilltop}) shows the form of the potential
  for the hilltop inflation.
 \begin{figure}[h]
 \centering
 \begin{subfigure}[b]{0.4\textwidth}
 \centering
\includegraphics[width=7cm]{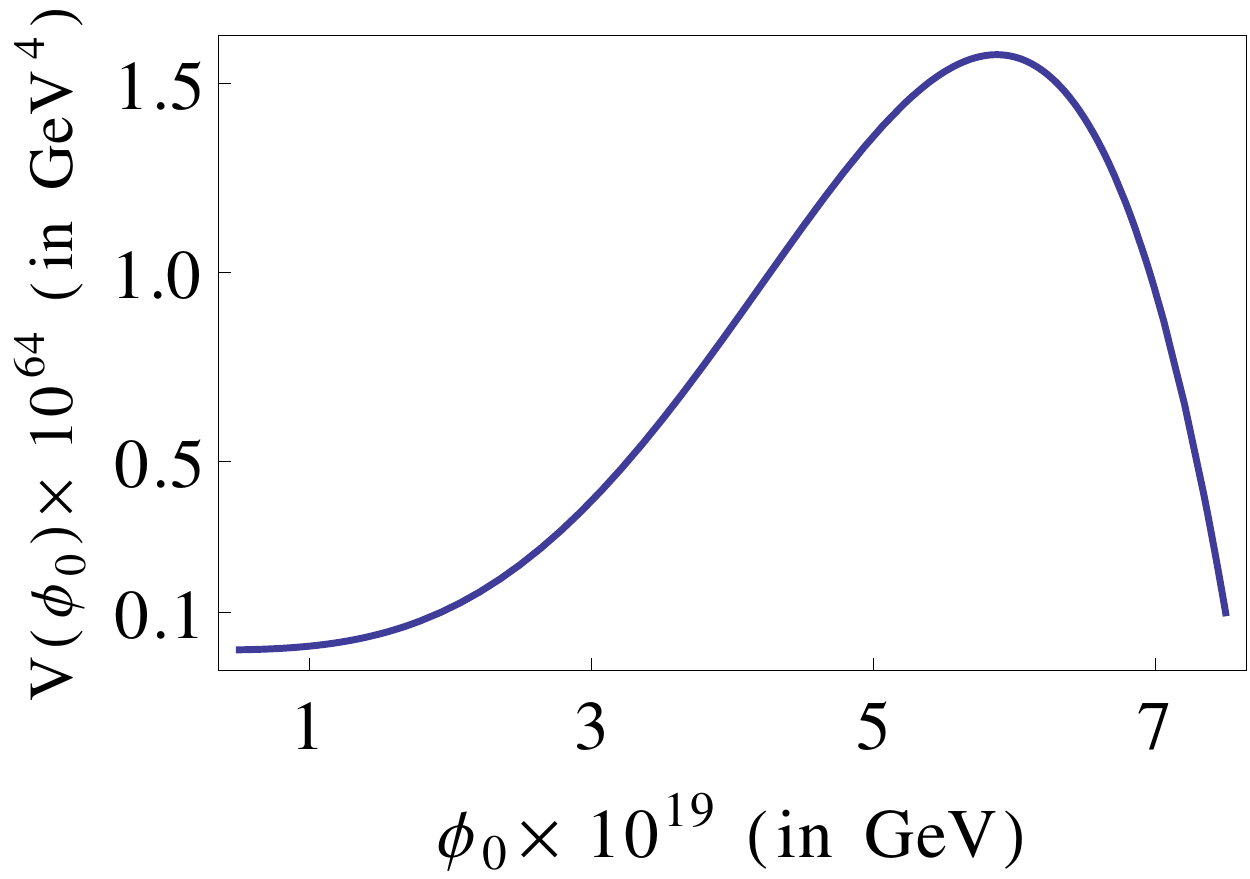}
\caption{Hilltop inflation potential}
\label{hilltop}
\end{subfigure}
\hfill
\begin{subfigure}[b]{0.5\textwidth}
\includegraphics[width=7cm]{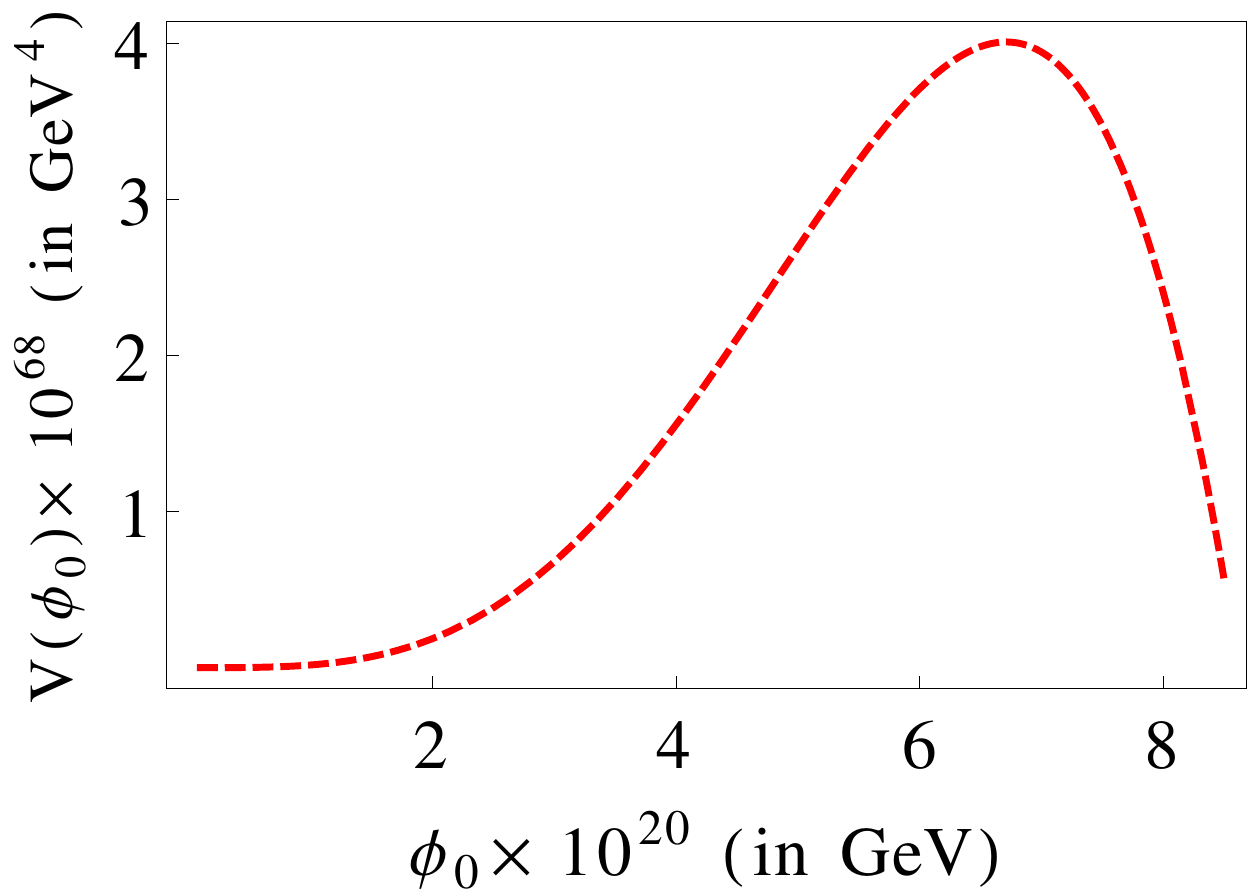}
\caption{$\phi^4-$branch inflation potential}
\label{phi4-pot}
\end{subfigure}
\end{figure}

\item {\bf $\phi^4-$branch inflation :} If one sets the scale of inflation to be $10^{16}$ GeV like the hilltop case, 
 one gets  $a_*\sim-0.022$ for $u_*=-11.001$. This indicates that the field
  value at the end of inflation, when $\epsilon_V\approx1$, would be
  $\phi_{0_{\rm end}}\sim2.6\,M_{\rm Pl}$. This $u_*$ yields
  the tensor-to-scalar ratio as $r_*=0.203$ and the inflaton quartic
  coupling, from the observation of the scalar power amplitude, as
  $\lambda_2\sim3.6\times10^{-13}$.  This provides the tree-level
  mass of the inflaton as $m_\phi\sim6.0\times10^9$ GeV. The running
  of the spectral index in such a scenario would be $\frac{dn_s}{d\ln
    k}|_{k_*}\sim -5.6\times10^{-4}$. In this scenario the
  inflaton-Higgs coupling can be of the order of $\sim10^{-6}$, which
  yields the reheating temperature as $T_{\rm RH}\sim1.09\times10^{13}$
  GeV. This reheating temperature and the energy-scale of inflation generate
  the $e$-folding at which pivot scale would have left the horizon as
  $N_*\sim67$. Fig.~(\ref{phi4-pot}) shows the form of the potential
  for the $\phi^4-$branch inflation. 
\end{itemize}

%==============================================================================
\section{Results}
\label{res}
In the previous section we have described the particle content of the
$U(1)_{B-L}$ extended SM, the dynamics of its scalar sector and the
framework of how inflation can be included in this model. Our aim is
to stabilize the vacuum all the way up to the Planck scale and also to
accommodate inflationary paradigm successfully. In this section we
would show how the aspect of vacuum stability can be achieved in our framework if
inflation is driven by the inflaton's radiatively corrected quartic self-interaction term.
%==============================================================================
%%%%%%%%%%%%%%%%%%%%%%%%%%%%%%%%%%%%%%%%%%%%%%%%%%%%%%%%%%%%%%%%%%%%%%%%%%%
\subsection{Vacuum Stability}

In the previous subsection, we have shown that to achieve successful
inflation, both the inflaton quartic coupling and the interaction
quartic coupling have to be fine tuned. Fine-tuning of inflaton
quartic coupling evidently brings down the mass scale of the inflaton
field which turns out to be below the instability scale of the
electroweak vacuum. Following \cite{EliasMiro:2012ay}, one can
integrate out the heavy inflaton field below its mass scale which then
adds a tree-level threshold correction to the low energy effective
Higgs quartic coupling $\lambda_S$ as (see Eqn.~(\ref{Veff}))
\begin{eqnarray}
\lambda_1=\lambda_S+\frac{\lambda_3^2}{4\lambda_2}.
\label{threshold}
\end{eqnarray}
Hence below the inflaton mass scale the stability condition
$(\lambda_S>0)$ for the SM Higgs quartic coupling would get shifted
upwards
$\lambda_1>\delta\lambda\equiv\frac{\lambda_3^2}{4\lambda_2}$. The
other two quartic couplings $\lambda_2$ and $\lambda_3$ would start
evolving at energies above this mass scale. The relevant RGEs are
written at the appendix \ref{app:quartic_RGEs}. To illustrate the
threshold effect, the running of Higgs quartic coupling has been shown
in Fig.~\ref{fig:Jul_15_lambda_all} for both the inflationary scenarios. 
Here we have taken the heavy
scalar mass $m_\phi=4\times10^9$ GeV before which $\lambda_S$ runs
according to SM $\beta$-functions. Beyond that point new loop effects
due to the extended theory starts to affect its running and the
discrete jump in the Higgs quartic coupling at $m_\phi$ is due to the
threshold effect.
 \begin{figure}[h]
\begin{center}
\includegraphics[width=9.0cm,angle=270]{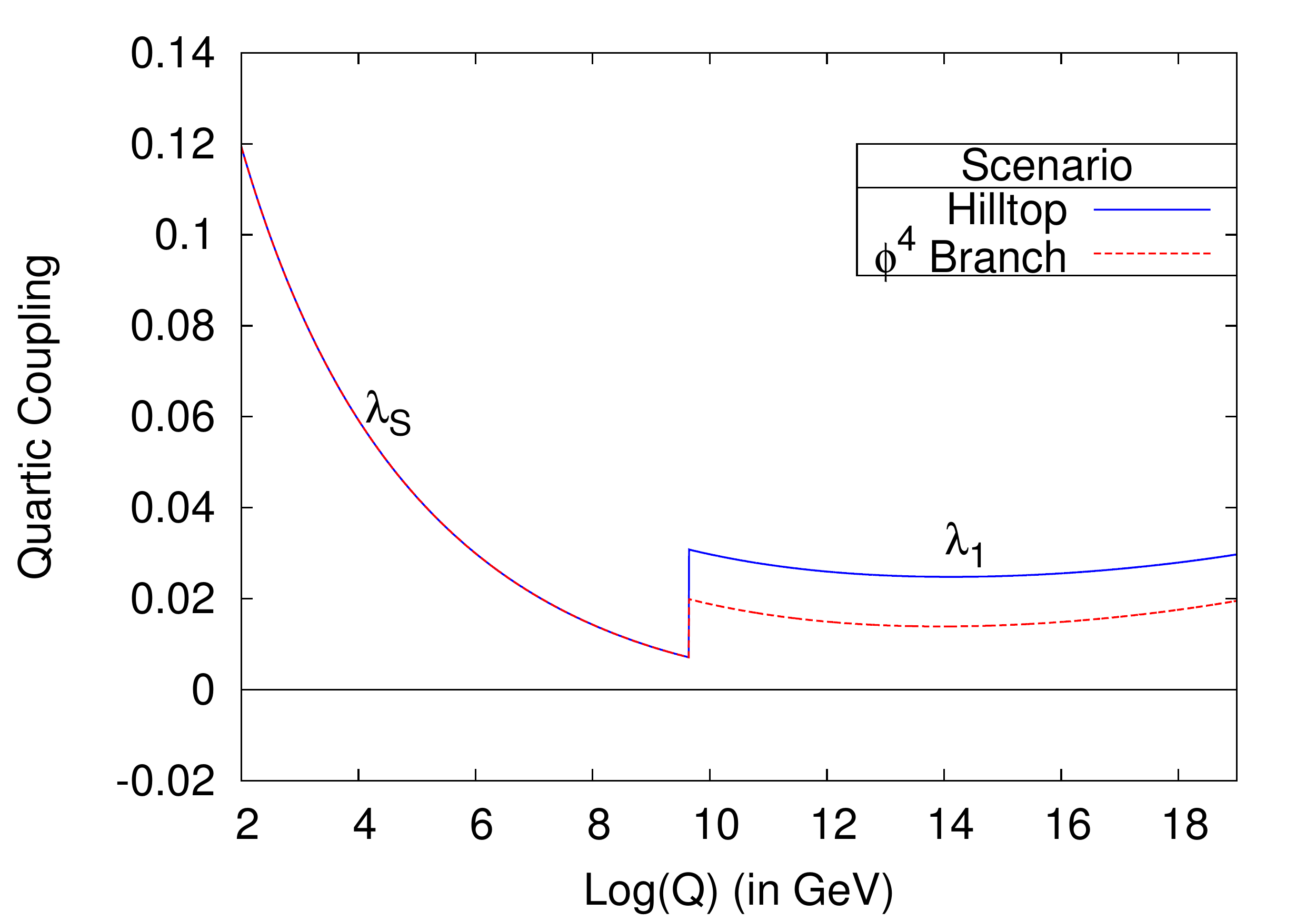}
\caption{This plot shows the running of the SM quartic coupling as a
  function of energy scale. The discrete jump at scale $\sim10^9$ GeV
  is because of the presence of the inflaton having mass $\sim 10^9$
  GeV. }
\label{fig:Jul_15_lambda_all}
\end{center}
\end{figure}

Apart from the SM fermions this model also contains three right handed
neutrinos, $N_{Ri}$, which appear in the Lagrangian as
\begin{eqnarray}
-{\mathcal L}_Y = Y^{\nu_L}_{ij} \bar{l}_{li} \tilde{S} N_{Rj}
+ Y^{N_R}_{ij}\, \overline{(N_R)^c_i}\,N_{Rj}\,\Phi + h.c.\,,
\end{eqnarray}
where $\tilde{S}=i\sigma_2 S^*$, $\sigma_2$ being the second Pauli
matrix and $l$ stands for the SM lepton doublet.  The second term in
the above Lagrangian gives rise to the coupling of the inflaton to
heavy right handed neutrinos and also masses for $N_{R}$. It is
important to note that when the $(B-L)$ symmetry is broken at the TeV
scale the masses of the right handed neutrinos are less compared to
the present scenario. In case of TeV scale breaking the Yukawa
couplings $(Y^{\nu_L})$ giving rise to the Dirac mass of light
neutrinos have to be vanishingly small unless some special textures
are considered. Thus in such cases, impact of $Y^{\nu_L}$ in the
evolutions of the quartic and other necessary couplings is
negligible. But in the present case the right handed neutrino masses
are very heavy $\sim 10^{11-13}$ GeV, due to high $U(1)_{B-L}$
breaking scale. Thus the light neutrino masses are still light $\sim
\mathcal{O}$(eV) even with $Y^{\nu_L} \sim\mathcal{O}(1)$. Hence
unlike the cases, where $U(1)_{B-L}$ symmetry is broken at TeV scale,
one can not ignore the contributions of light neutrino Yukawa
couplings $Y^{\nu_L}$ in the RGEs in our scenario (see
appendix~\ref{app:Yukawa_RGEs}).

%===========================================================================
\subsection{Other constraints}

Looking at the threshold correction, given in Eqn.~(\ref{threshold}),
which is essential for electroweak vacuum stability, it may seem that
$\lambda_3 < 0$ can still be retained as a possible condition. But, in
our analysis this opportunity of achieving larger parameter space for
$\lambda_3$ is restricted as here $\lambda_2$ is very small $\sim
10^{-14}$ due to inflationary constraints. The absolute value of
$\lambda_3$ can never be too large as it affects the running of
$\lambda_2$ by driving its value to a much larger value which might
not be able to explain inflationary dynamics.  Thus $\lambda_3$ is
constrained from above by the requirement of inflation. The smallness
of $|\lambda_3|$ ensures that the two scalars present in the theory
are basically decoupled from each other as the mixing angle between
then becomes too small, see Eqn.~(\ref{mixing-angle}). This confirms
that the `decoupling theorem' holds good in our scenario.

At the inflaton mass scale from which running of $\lambda_2$ and
$\lambda_3$ become important, explicit values of $\lambda_3$ and
$\lambda_2$ have been set as $8.0\times10^{-8}$ and $5\times 10^{-13}$ ($2.7\times10^{-13}$) for hilltop inflation ($\phi^4$-branch inflation) 
respectively. These values ensures both vacuum stability up to the
Planck scale and inflation with correct scalar amplitude at GUT
scale. The $B-L$ gauge coupling, $g_{B-L}$, also affects the running
of $\lambda_2$ from $m_{\phi}$ scale to inflation scale (see
Appendix.~(\ref{app_rg})). The radiative correction to the scalar
potential is shown in Eqn.~(\ref{eqn:radiative-correction}) which
depends on both $\lambda_3$ and $g_{B-L}$. We set $g_{B-L}$ at $\sim
10^{-5}$  and show the variation of $a$ with energy scale in Fig.~\ref{fig:a-param} for both the inflationary scenarios.
 \begin{figure}[t!]
\begin{center}
\includegraphics[width=10.0cm,angle=270]{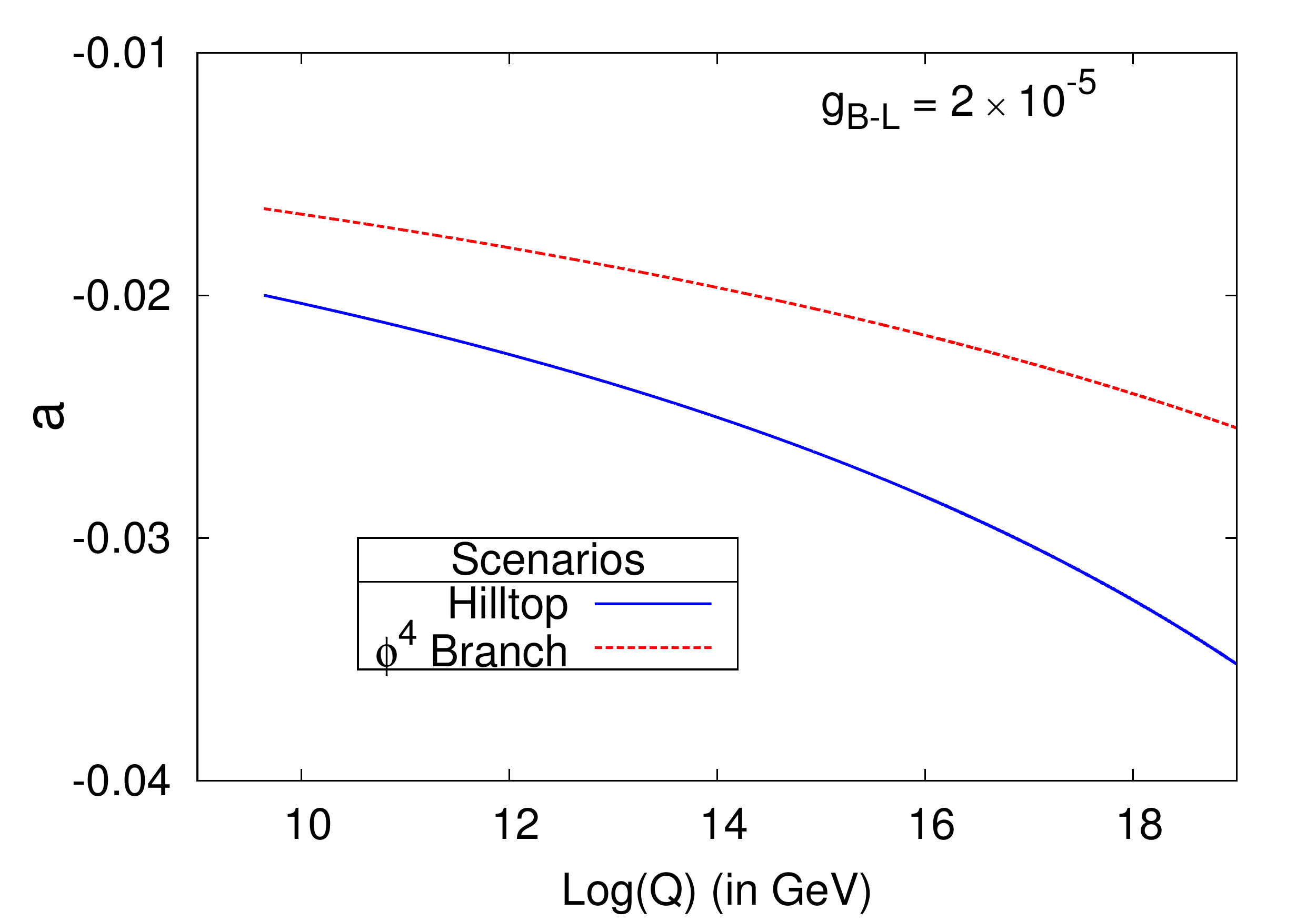}
\caption{Variation of $a$, see Eqn.~(\ref{eqn:radiative-correction}),
  with respect to the energy scale. }
\label{fig:a-param}
\end{center}
\end{figure}
%============================================================================

In a previous discussion we have presented a possible mode in which
the inflaton can decay into a pair of Higgs fields during
reheating. In passing, we note another possible way of reheating in
our model.  The inflaton can also decay to two heavy right handed
neutrinos where the decay rate is given by
\begin{eqnarray} 
\Gamma_N(\phi_0 \to N_{Ri} N_{Ri} ) = \frac{\left(Y^{N_R}_i\right)^2}{64\pi}\,m_\phi\,. 
\label{gamf}
\end{eqnarray}
But this $\Gamma_N$ is much smaller than $\Gamma_S(\phi_0\rightarrow
SS)$ (as $Y_i^{N_R}\sim10^{-4}$ in our scenario) yielding a much
smaller reheating temperature compared to $T_{\rm RH}$ when the
inflaton prominently decays into Higgs. In general if both the decay
channels are open for the inflaton, reheating will prominently happen
through the Higgs decay of the inflaton.
%===========================================================================
\subsection{Quadratic divergence of Higgs mass and the Veltman criteria}

We would like to note in passing that the stabilization of the SM
Higgs boson mass under the quadratic divergences specially in high
scale theory is an unavoidable issue. The generic problem with any
high scale non-supersymmetric models is related to the stabilization
of scalar masses, specially the SM Higgs mass. Due to the quadratic
divergences the SM Higgs mass acquires a correction proportional to
$\Lambda^2$, where $\Lambda$ is the scale of new physics. In
supersymmetric theory these corrections automatically get cancelled
out with the ones coming from their supersymmetric partners. To avoid
such large contributions to scalar mass in non-supersymmetric models
one needs to impose the Veltman condition. This prescription, as
suggested in \cite{Veltman:1980mj,Einhorn:1992um}, confirms the
removal of quadratic divergences of the scalar masses to stabilize
them. We note here that in our model the scalar masses might obtain a
large radiative correction.  To avoid such catastrophe in scalar
masses we need to satisfy the Veltman criteria (VC), which for this
scenario would be \cite{Khalil:2006yi} \bea \delta m_s^2 & \propto &
v_{\phi}^2 [ (2\lambda_1 +\lambda_3/3) \cos^4 \alpha
  +\left(\frac{2M_W^2+M_Z^2}{v^2}\right) \cos^2 \alpha \nonumber \\ &
  & + 4 g_{B-L}^2 \sin^2 \alpha -4(Y_t^2 \cos^2 \alpha + (Y^{\nu_L
  })^2 \cos^2 \alpha + (Y^{\nu_R})^2 \sin^2 \alpha) ], \eea with
$\cos^2 \alpha \sim 0.99879$, where $\alpha$ is the mixing angle as
given in Eqn.~(\ref{mixing-angle}).  Since $v_{\phi}$ is very large
for our case, $Z_{B-L}$ and $\nu_R$ will not affect this criteria
much. Also $\lambda_3$ is very small. But the light neutrino Yukawa
coupling can be large here and thus its impact can be sizeable. This
has been added with top quark contribution.  We find that within the
available parameter space in our model it is indeed possible to
satisfy VC either at the inflation scale or at Planck scale. With
light neutrino Yukawa to be 0.1462 and  0.2413 the VC can be satisfied
at the inflation and the Planck scales respectively.  This implies
that the light Higgs remains light at the inflation scale and the
decay of the inflaton, considered in this paper for explaining the
reheating, does not suffer any catastrophe under the impact of
radiative corrections. Thus we can stabilize the Higgs mass at the
inflation scale but perhaps this mechanism does not solve the
stabilization at other scales.

%%%%%%%%%%%%%%%%%%%%%%%%%%%%%%%%%%%%%%%%%%%%%%%%%%%%%%%%%%%%
\section{Discussion and conclusion}
\label{concl}
The recent discovery of SM Higgs scalar and the detection of B-mode
polarization of CMBR by BICEP2, if stands the test of time, are quite remarkable as these
observations would help understand the evolution of our
universe. Though these two observations carry signatures of physics
taking place at very different scales, we tried to connect them within
a single particle physics picture in this article. The observation by
BICEP2 of the CMBR B-mode polarization, if confirmed, would also
indicate the scale of inflation to be around $\sim 10^{16}$
GeV. This energy scale is quite interesting in the field of particle
physics as incidentally this scale turns out to be the unification
scale. In this work, we have adopted a gauge extended SM scenario
which contains a SM singlet scalar field with a new $U(1)_{B-L}$ gauge
charge. This field acquires vev at a very high scale and breaks the
$U(1)_{B-L}$ symmetry spontaneously and also couples to the SM
particles.  Apart from this SM singlet scalar, there are three right
handed neutrinos which successfully generates the light neutrino
masses through type-I seesaw without fine-tuning the Dirac Yukawa
coupling.

The electroweak breaking scale is around $246$ GeV whereas in this
case the $U(1)_{B-L}$ breaking scale should lie near the GUT scale so
that such high-scale inflation can take place as has been demanded by
BICEP2. But, as the $U(1)_{B-L}$ and electroweak breaking scales lie
far apart in our scenario, these two theories are basically decoupled
from each other, as has been demanded by the `decoupling
theorem'. Then it might imply that extending the SM by such high scale
$U(1)_{B-L}$ gauge theory fails to serve its purpose of taking care of
the stability of electroweak vacuum. But the advantage of introducing
such high scale $U(1)_{B-L}$ symmetry is that it provides a heavy
scalar $\Phi$ in the theory, whose real part plays the role of the
inflaton in such a scenario. Presence of a heavy scalar yields a
threshold correction to the Higgs quartic coupling, if one integrates
out this heavy scalar below its mass scale. Hence if the mass of this
heavy scalar lies below the electroweak instability scale
($\sim10^{10}$ GeV), the threshold correction eventually helps
avoid the instability of the vacuum by correctly uplifting the value
of the SM Higgs quartic coupling at this mass scale. Hence the key
point is to keep the mass scale of the heavy scalar in the theory
below the electroweak scale if one wants the threshold corrections to
help stabilize the vacuum, even though the explicit value of the mass
of the heavy scalar does not play any important role.
 
Even at this point, it might seem that the heavy scalar mass, which is
basically the inflaton mass in our case, would lie above the
electroweak instability scale as the $U(1)_{B-L}$ breaking scale lies
far above this instability scale. Thus it might seem that even the
threshold effect would not help to keep the vacuum stable in such a
case. The important point to note here is that the heavy scalar, which
yields the threshold correction to the SM Higgs quartic coupling, also
plays the role of the inflaton in such a case. In the large field
inflation scenario, supported by BICEP2 observations, inflation is
naturally driven by the radiatively corrected quartic potential of the inflaton. Thus the
inflaton's quartic coupling parameter, which eventually determines its
mass, has to be extremely fine-tuned so that the amplitude of the
scalar power spectrum remains in accordance with observation. Such
high fine-tuning of inflaton quartic coupling evidently brings down
the mass of the inflaton below the electroweak vacuum instability
scale, enabling the threshold effect to keep the electroweak vacuum
stable. 

The radiatively corrected inflation potential can also produce much lower tensor-to-scalar 
$r$ in the hilltop branch inflation scenario, which remains much below the upper-limit set by PLANCK and hence observationally viable.
The other scenario, namely $\phi^4-$branch inflation,  produces $r\sim0.203$ which is compatible with BICEP2 observations. Both these 
 scenarios produce very low running of the scalar spectral index $(\mathcal O(10^{-4}))$. Thus the $\phi^4-$branch inflation scenario would not
 resolve the discrepancy between PLANCK and BICEP2 observation of $r$, while the hilltop inflationary scenario is observationally safe if the BICEP2 observations are
 ruled out in future. 

With such a set-up, we have computed the correlated bounds on scalar
quartic couplings and Yukawa couplings compatible with vacuum
stability and perturabativity all the way up to the Planck scale. In
order to have successful inflation in this particle physics model, we
showed that fine-tuning of $U(1)_{B-L}$ quartic couplings allow a
chaotic-type inflation driven by the radiatively corrected quartic potential of the real
part of the SM singlet scalar and the scenario remains in accordance
with the observations of BICEP2 and PLANCK. We are also able to
ascertain a particular parameter space where the criteria for vacuum
stability and requirements to explain inflationary dynamics are
simultaneously satisfied. To our advantage, this model also yields
reheating after inflation through the channel where the inflaton
possibly decays into a pair of SM Higgs field.  Constraints coming
from such reheating scenario are automatically satisfied within our
choice of parameter space. Hence to summarize, the $U(1)_{B-L}$
symmetry extended SM considered here is capable of
\begin{itemize}
\item stabilizing SM electro-weak vacuum all the way up to the Planck scale,
\item yielding light neutrino masses without fine-tuning of Dirac
  Yukawa couplings,
\item accommodate large-field chaotic-type inflation which is in
  accordance with BICEP2 and PLANCK's observations,
\item successfully reheat the universe at the end of inflation.
\end{itemize}

%%%%%%%%%%%%%%%%%%%%%%%%%%%%%%%%%%%%%%%%%%%%%%%%%%%%%%%%%%%%%%%%%%%%%%
\begin{acknowledgments}
 Works of JC and SD are supported by Department of Science and
 Technology, Government of INDIA under the Grant Agreement numbers
 IFA12-PH-34 (INSPIRE Faculty Award) and IFA13-PH-77 (INSPIRE Faculty
 Award) respectively.
\end{acknowledgments}

%==============================================================================
\appendix

\section{Renormalisation  of Couplings}
\label{app_rg}
Here we have depicted the necessary renormalisation group (RG) evolutions for $U(1)_{B-L}$ extended Standard Model as suggested in  \cite{Basso:2010jm_U1BL, u1b-l_recent, Basso:2013vla}. The necessary parameters  are: scalar quartic $(\lambda_{1,2,3})$, top Yukawa $(Y_t)$, Dirac Yukawa  
$(Y^{\nu_L})$, right handed neutrino Yukawa $(Y^{N_R})$, gauge $(g_{1,2},g_{B-L},g^{'})$ couplings.

\subsection{ RG equations for scalar quartic couplings:}\label{app:quartic_RGEs}

Here we have illustrated the RGEs for $\lambda_{1,2,3}$. Their evolutions with scales involve the renormalisation of other parameters, see Eqns.~\ref{app:Yukawa_RGEs}, and \ref{app:gauge_RGEs}.
\bea
\left(16 \pi^2\right)\; \frac{d\lambda_1}{dt} &=&  \Bigg[ 24 \lambda_1^2 + \lambda_3^2 - 6 Y_t^4 +\frac{9}{8}g_2^4 + 
						    \frac{3}{8}g_1^4+\frac{3}{4}g_1^2 g_2^2    + \frac{3}{4}g_1^2 g^{'2} + 
						    \frac{3}{4}g_2^2 g^{'2} \nonumber \\
					       & & + \frac{3}{4}g^{'4} + 9 \lambda_1 g_2^2 + 12\lambda_1 Y_t^2  - 3 \lambda_1 g_1^2  
						  - 3 \lambda_1 g^{'2} -2 (Y^{\nu_L})^4 + 4\lambda_1 (Y^{\nu_L})^2 \Bigg], \nonumber \\
\left(16 \pi^2\right)\; \frac{d\lambda_2}{dt} &=&  \Bigg[20\lambda_2^2+2\lambda_3^2-
\textrm{Tr}[(Y^{N_R})^4]+ 96 g_{B-L}^4 +   8 \lambda_2^2\; \textrm{Tr}[(Y^{N_R})^2]i.e.,
						    -48 \lambda_2 g_{B-L}^2\Bigg], \\     
\left(16 \pi^2\right)\; \frac{d\lambda_3}{dt} &=&  \lambda_3 \Bigg[ 12 \lambda_1 + 8 \lambda_2 + 4 \lambda_3 +6 Y_t^2 - \frac{3}{2}(3g_2^2 - g_1^2-g^{'2})
						   \nonumber \\ 
					       & & + 4\textrm{Tr}[(Y^{N_R})^2] -24 g_{B-L}^2  + 12 g_{B-L}^2 g^{'2}\Bigg]. \nonumber
\eea

\subsection{ RG equations for Yukawa couplings:}\label{app:Yukawa_RGEs}

Here we have characterized the RGEs of different Yukawa couplings that affect the evolutions of quartic couplings as given in Eqn.~\ref{app:quartic_RGEs}.
\bea
\left(16 \pi^2\right)\; \frac{d Y_t}{dt} &=&  Y_t \Bigg[ \frac{9}{2} Y_t^2 -8g_3^2 -\frac{9}{4}g_2^2 -\frac{17}{12}(g_1^2+g_2^2+g^{'2})-\frac{2}{3}g_{B-L}^2 -\frac{5}{3}g_{B-L}g^{'}+(Y^{\nu_L})^2\Bigg], \nonumber \\
\left(16 \pi^2\right)\; \frac{d Y^{N_R}}{dt} &=&  Y^{N_R} \Bigg[4 (Y^{N_R})^2 + 2\; \textrm{Tr} [(Y^{N_R})^2] 
- 6 g_{B-L}^2 \Bigg],\\
\left(16 \pi^2\right)\; \frac{d Y^{\nu_L}}{dt} &=& Y^{\nu_L} \Bigg[\frac{5}{2}(Y^{\nu_L})^2+3Y_t^2-\frac{9}{4}g_2^2-\frac{3}{4}g_1^2-6 g_{B-L}^2 \Bigg]. \nonumber
\eea

\subsection{ RG equations for the gauge couplings:}\label{app:gauge_RGEs}

The gauge coupling runnings are delineated here as they also play crucial role while adjudging the RGEs of quartic couplings and Yukawa couplings, see Eqns.~\ref{app:quartic_RGEs}, and \ref{app:Yukawa_RGEs}.
\bea
\left(16 \pi^2\right)\;\frac{d g_3}{dt} &=&  \Big[-7\Big]  g_3^3, \nonumber \\
\left(16 \pi^2\right)\;\frac{d g_2}{dt} &=&  \Big[-\frac{19}{6}\Big]  g_2^3, \nonumber \\
\left(16 \pi^2\right)\;\frac{d g_1}{dt} &=&  \Big[\frac{41}{6}\Big]  g_1^3, \\
\left(16 \pi^2\right)\;\frac{d g_{B-L}}{dt} &=&  \Bigg[12 g_{B-L}^3 + (32/3) g_{B-L}^2 g' +(41/6) g_{B-L} g'^{2}\Bigg], \nonumber \\
\left(16 \pi^2\right)\;\frac{d g^{'}}{dt} &=&  \Bigg[\frac{41}{6} (g'^{3}+2g_1^2 g') + \frac{32}{3} g_{B-L} (g'^{2}+g_1^2) + 12 g_{B-L}^2 g'\Bigg]. \nonumber
\eea

%------------------------------------------------------------------------------
\providecommand{\href}[2]{#2}
\addcontentsline{toc}{section}{References}
\bibliographystyle{JHEP}
\bibliography{gut-inf-bib}

\providecommand{\href}[2]{#2}\begingroup\raggedright\begin{thebibliography}{10}

\bibitem{Aad:2012tfa}
{\bf ATLAS Collaboration} Collaboration, G.~Aad et~al., {\it {Observation of a
  new particle in the search for the Standard Model Higgs boson with the ATLAS
  detector at the LHC}},  {\em Phys.Lett.} {\bf B716} (2012) 1--29,
  [\href{http://xxx.lanl.gov/abs/1207.7214}{{\tt arXiv:1207.7214}}].

\bibitem{Chatrchyan:2012ufa}
{\bf CMS Collaboration} Collaboration, S.~Chatrchyan et~al., {\it {Observation
  of a new boson at a mass of 125 GeV with the CMS experiment at the LHC}},
  {\em Phys.Lett.} {\bf B716} (2012) 30--61,
  [\href{http://xxx.lanl.gov/abs/1207.7235}{{\tt arXiv:1207.7235}}].

\bibitem{Ade:2014xna}
{\bf BICEP2 Collaboration} Collaboration, P.~Ade et~al., {\it {BICEP2 I:
  Detection Of B-mode Polarization at Degree Angular Scales}},
  \href{http://xxx.lanl.gov/abs/1403.3985}{{\tt arXiv:1403.3985}}.

\bibitem{Coleman:1977py_sm_old}
S.~R. Coleman, {\it {The Fate of the False Vacuum. 1. Semiclassical Theory}},
  {\em Phys.Rev.} {\bf D15} (1977) 2929--2936.

\bibitem{Sher:1988mj}
M.~Sher, {\it {Electroweak Higgs Potentials and Vacuum Stability}},  {\em
  Phys.Rept.} {\bf 179} (1989) 273--418.

\bibitem{Sher:1993mf_sm_old}
M.~Sher, {\it {Precise vacuum stability bound in the standard model}},  {\em
  Phys.Lett.} {\bf B317} (1993) 159--163,
  [\href{http://xxx.lanl.gov/abs/hep-ph/9307342}{{\tt hep-ph/9307342}}].

\bibitem{Espinosa:1995se_sm_old}
J.~Espinosa and M.~Quiros, {\it {Improved metastability bounds on the standard
  model Higgs mass}},  {\em Phys.Lett.} {\bf B353} (1995) 257--266,
  [\href{http://xxx.lanl.gov/abs/hep-ph/9504241}{{\tt hep-ph/9504241}}].

\bibitem{Schrempp:1996fb_sm_old}
B.~Schrempp and M.~Wimmer, {\it {Top quark and Higgs boson masses: Interplay
  between infrared and ultraviolet physics}},  {\em Prog.Part.Nucl.Phys.} {\bf
  37} (1996) 1--90, [\href{http://xxx.lanl.gov/abs/hep-ph/9606386}{{\tt
  hep-ph/9606386}}].

\bibitem{Casas:1996aq_sm_old}
J.~Casas, J.~Espinosa, and M.~Quiros, {\it {Standard model stability bounds for
  new physics within LHC reach}},  {\em Phys.Lett.} {\bf B382} (1996) 374--382,
  [\href{http://xxx.lanl.gov/abs/hep-ph/9603227}{{\tt hep-ph/9603227}}].

\bibitem{Isidori:2001bm_sm_old}
G.~Isidori, G.~Ridolfi, and A.~Strumia, {\it {On the metastability of the
  standard model vacuum}},  {\em Nucl.Phys.} {\bf B609} (2001) 387--409,
  [\href{http://xxx.lanl.gov/abs/hep-ph/0104016}{{\tt hep-ph/0104016}}].

\bibitem{Holthausen:2011aa_smRGE}
M.~Holthausen, K.~S. Lim, and M.~Lindner, {\it {Planck scale Boundary
  Conditions and the Higgs Mass}},  {\em JHEP} {\bf 1202} (2012) 037,
  [\href{http://xxx.lanl.gov/abs/1112.2415}{{\tt arXiv:1112.2415}}].

\bibitem{Alekhin:2012py_vs}
S.~Alekhin, A.~Djouadi, and S.~Moch, {\it {The top quark and Higgs boson masses
  and the stability of the electroweak vacuum}},  {\em Phys.Lett.} {\bf B716}
  (2012) 214--219, [\href{http://xxx.lanl.gov/abs/1207.0980}{{\tt
  arXiv:1207.0980}}].

\bibitem{Degrassi:2012ry_vs}
G.~Degrassi, S.~Di~Vita, J.~Elias-Miro, J.~R. Espinosa, G.~F. Giudice, et~al.,
  {\it {Higgs mass and vacuum stability in the Standard Model at NNLO}},  {\em
  JHEP} {\bf 1208} (2012) 098, [\href{http://xxx.lanl.gov/abs/1205.6497}{{\tt
  arXiv:1205.6497}}].

\bibitem{Masina:2012tz_vs}
I.~Masina, {\it {The Higgs boson and Top quark masses as tests of Electroweak
  Vacuum Stability}},  {\em Phys.Rev.} {\bf D87} (2013) 053001,
  [\href{http://xxx.lanl.gov/abs/1209.0393}{{\tt arXiv:1209.0393}}].

\bibitem{Tang:2013bz_sm}
Y.~Tang, {\it {Vacuum Stability in the Standard Model}},  {\em Mod.Phys.Lett.}
  {\bf A28} (2013) 1330002, [\href{http://xxx.lanl.gov/abs/1301.5812}{{\tt
  arXiv:1301.5812}}].

\bibitem{Herranen:2014cua}
M.~Herranen, T.~Markkanen, S.~Nurmi, and A.~Rajantie, {\it {Spacetime curvature
  and the Higgs stability during inflation}},
  \href{http://xxx.lanl.gov/abs/1407.3141}{{\tt arXiv:1407.3141}}.

\bibitem{Spencer-Smith:2014woa}
A.~Spencer-Smith, {\it {Higgs Vacuum Stability in a Mass-Dependent
  Renormalisation Scheme}},  \href{http://xxx.lanl.gov/abs/1405.1975}{{\tt
  arXiv:1405.1975}}.

\bibitem{Gabrielli:2013hma}
E.~Gabrielli, M.~Heikinheimo, K.~Kannike, A.~Racioppi, M.~Raidal, et~al., {\it
  {Towards Completing the Standard Model: Vacuum Stability, EWSB and Dark
  Matter}},  {\em Phys.Rev.} {\bf D89} (2014) 015017,
  [\href{http://xxx.lanl.gov/abs/1309.6632}{{\tt arXiv:1309.6632}}].

\bibitem{Branchina:2013jra}
V.~Branchina and E.~Messina, {\it {Stability, Higgs Boson Mass and New
  Physics}},  {\em Phys.Rev.Lett.} {\bf 111} (2013) 241801,
  [\href{http://xxx.lanl.gov/abs/1307.5193}{{\tt arXiv:1307.5193}}].

\bibitem{Branchina:2014rva}
V.~Branchina, E.~Messina, and M.~Sher, {\it {The lifetime of the electroweak
  vacuum and sensitivity to Planck scale physics}},
  \href{http://xxx.lanl.gov/abs/1408.5302}{{\tt arXiv:1408.5302}}.

\bibitem{Branchina:2014usa}
V.~Branchina, E.~Messina, and A.~Platania, {\it {Top mass determination, Higgs
  inflation, and vacuum stability}},
  \href{http://xxx.lanl.gov/abs/1407.4112}{{\tt arXiv:1407.4112}}.

\bibitem{Kobakhidze:2014xda}
A.~Kobakhidze and A.~Spencer-Smith, {\it {The Higgs vacuum is unstable}},
  \href{http://xxx.lanl.gov/abs/1404.4709}{{\tt arXiv:1404.4709}}.

\bibitem{Casas:1994qy_sm_old}
J.~Casas, J.~Espinosa, and M.~Quiros, {\it {Improved Higgs mass stability bound
  in the standard model and implications for supersymmetry}},  {\em Phys.Lett.}
  {\bf B342} (1995) 171--179,
  [\href{http://xxx.lanl.gov/abs/hep-ph/9409458}{{\tt hep-ph/9409458}}].

\bibitem{Anchordoqui:2012fq_sm++}
L.~A. Anchordoqui, I.~Antoniadis, H.~Goldberg, X.~Huang, D.~Lust, et~al., {\it
  {Vacuum Stability of Standard Model$^{++}$}},  {\em JHEP} {\bf 1302} (2013)
  074, [\href{http://xxx.lanl.gov/abs/1208.2821}{{\tt arXiv:1208.2821}}].

\bibitem{Hung:1979dn_sm_old}
P.~Q. Hung, {\it {Vacuum Instability and New Constraints on Fermion Masses}},
  {\em Phys.Rev.Lett.} {\bf 42} (1979) 873.

\bibitem{Chakrabortty:2013zja}
J.~Chakrabortty, P.~Konar, and T.~Mondal, {\it {Constraining a class of B-L
  extended models from vacuum stability and perturbativity}},  {\em Phys.Rev.}
  {\bf D89} (2014) 056014, [\href{http://xxx.lanl.gov/abs/1308.1291}{{\tt
  arXiv:1308.1291}}].

\bibitem{u1b-l_recent}
A.~Datta, A.~Elsayed, S.~Khalil, and A.~Moursy, {\it {Higgs Vacuum Stability in
  $B-L$ extended Standard Model}},
  \href{http://xxx.lanl.gov/abs/1308.0816}{{\tt arXiv:1308.0816}}.

\bibitem{Chakrabortty:2013mha}
J.~Chakrabortty, P.~Konar, and T.~Mondal, {\it {Copositive Criteria and
  Boundedness of the Scalar Potential}},
  \href{http://xxx.lanl.gov/abs/1311.5666}{{\tt arXiv:1311.5666}}.

\bibitem{Guth:1980zm}
A.~H. Guth, {\it {The Inflationary Universe: A Possible Solution to the Horizon
  and Flatness Problems}},  {\em Phys.Rev.} {\bf D23} (1981) 347--356.

\bibitem{Linde:1983gd}
A.~D. Linde, {\it {Chaotic Inflation}},  {\em Phys.Lett.} {\bf B129} (1983)
  177--181.

\bibitem{Ade:2013zuv}
{\bf Planck Collaboration} Collaboration, P.~Ade et~al., {\it {Planck 2013
  results. XVI. Cosmological parameters}},
  \href{http://xxx.lanl.gov/abs/1303.5076}{{\tt arXiv:1303.5076}}.

\bibitem{Adam:2014bub}
{\bf Planck Collaboration} Collaboration, R.~Adam et~al., {\it {Planck
  intermediate results. XXX. The angular power spectrum of polarized dust
  emission at intermediate and high Galactic latitudes}},
  \href{http://xxx.lanl.gov/abs/1409.5738}{{\tt arXiv:1409.5738}}.

\bibitem{Okada:2014lxa}
N.~Okada, V.~Nefer, and Q.~Shafi, {\it {Simple Inflationary Models in Light of
  BICEP2: an Update}},  \href{http://xxx.lanl.gov/abs/1403.6403}{{\tt
  arXiv:1403.6403}}.

\bibitem{Kobayashi:2014jga}
T.~Kobayashi and O.~Seto, {\it {Polynomial inflation models after BICEP2}},
  {\em Phys.Rev.} {\bf D89} (2014) 103524,
  [\href{http://xxx.lanl.gov/abs/1403.5055}{{\tt arXiv:1403.5055}}].

\bibitem{Bezrukov:2007ep}
F.~L. Bezrukov and M.~Shaposhnikov, {\it {The Standard Model Higgs boson as the
  inflaton}},  {\em Phys.Lett.} {\bf B659} (2008) 703--706,
  [\href{http://xxx.lanl.gov/abs/0710.3755}{{\tt arXiv:0710.3755}}].

\bibitem{Cook:2014dga}
J.~L. Cook, L.~M. Krauss, A.~J. Long, and S.~Sabharwal, {\it {Is Higgs
  Inflation Dead?}},  \href{http://xxx.lanl.gov/abs/1403.4971}{{\tt
  arXiv:1403.4971}}.

\bibitem{Fairbairn:2014nxa}
M.~Fairbairn, P.~Grothaus, and R.~Hogan, {\it {The Problem with False Vacuum
  Higgs Inflation}},  \href{http://xxx.lanl.gov/abs/1403.7483}{{\tt
  arXiv:1403.7483}}.

\bibitem{Hamada:2014iga}
Y.~Hamada, H.~Kawai, K.-y. Oda, and S.~C. Park, {\it {Higgs inflation still
  alive}},  \href{http://xxx.lanl.gov/abs/1403.5043}{{\tt arXiv:1403.5043}}.

\bibitem{Masina:2014yga}
I.~Masina, {\it {The Gravitational Wave Background and Higgs False Vacuum
  Inflation}},  \href{http://xxx.lanl.gov/abs/1403.5244}{{\tt
  arXiv:1403.5244}}.

\bibitem{Bezrukov:2014bra}
F.~Bezrukov and M.~Shaposhnikov, {\it {Higgs inflation at the critical point}},
   \href{http://xxx.lanl.gov/abs/1403.6078}{{\tt arXiv:1403.6078}}.

\bibitem{Nakayama:2014koa}
K.~Nakayama and F.~Takahashi, {\it {Higgs Chaotic Inflation and the Primordial
  B-mode Polarization Discovered by BICEP2}},
  \href{http://xxx.lanl.gov/abs/1403.4132}{{\tt arXiv:1403.4132}}.

\bibitem{Germani:2014hqa}
C.~Germani, Y.~Watanabe, and N.~Wintergerst, {\it {Self-unitarization of New
  Higgs Inflation and compatibility with Planck and BICEP2 data}},
  \href{http://xxx.lanl.gov/abs/1403.5766}{{\tt arXiv:1403.5766}}.

\bibitem{Oda:2014rpa}
I.~Oda and T.~Tomoyose, {\it {Quadratic Chaotic Inflation from Higgs
  Inflation}},  \href{http://xxx.lanl.gov/abs/1404.1538}{{\tt
  arXiv:1404.1538}}.

\bibitem{Feng:2014naa}
C.-J. Feng and X.-Z. Li, {\it {Is Cosmological Constant Needed in Higgs
  Inflation?}},  \href{http://xxx.lanl.gov/abs/1404.3817}{{\tt
  arXiv:1404.3817}}.

\bibitem{Haba:2014zda}
N.~Haba and R.~Takahashi, {\it {Higgs inflation with singlet scalar dark matter
  and right-handed neutrino in the light of BICEP2}},
  \href{http://xxx.lanl.gov/abs/1404.4737}{{\tt arXiv:1404.4737}}.

\bibitem{Fairbairn:2014zia}
M.~Fairbairn and R.~Hogan, {\it {Electroweak Vacuum Stability in light of
  BICEP2}},  {\em Phys.Rev.Lett.} {\bf 112} (2014) 201801,
  [\href{http://xxx.lanl.gov/abs/1403.6786}{{\tt arXiv:1403.6786}}].

\bibitem{Lebedev:2012sy}
O.~Lebedev and A.~Westphal, {\it {Metastable Electroweak Vacuum: Implications
  for Inflation}},  {\em Phys.Lett.} {\bf B719} (2013) 415--418,
  [\href{http://xxx.lanl.gov/abs/1210.6987}{{\tt arXiv:1210.6987}}].

\bibitem{Hamada:2012bp}
Y.~Hamada, H.~Kawai, and K.-y. Oda, {\it {Bare Higgs mass at Planck scale}},
  {\em Phys.Rev.} {\bf D87} (2013), no.~5 053009,
  [\href{http://xxx.lanl.gov/abs/1210.2538}{{\tt arXiv:1210.2538}}].

\bibitem{Marshak:1979fm_BL}
R.~Marshak and R.~N. Mohapatra, {\it {Quark - Lepton Symmetry and B-L as the
  U(1) Generator of the Electroweak Symmetry Group}},  {\em Phys.Lett.} {\bf
  B91} (1980) 222--224.

\bibitem{Buchmuller:1991ce_U1BL}
W.~Buchmuller, C.~Greub, and P.~Minkowski, {\it {Neutrino masses, neutral
  vector bosons and the scale of B-L breaking}},  {\em Phys.Lett.} {\bf B267}
  (1991) 395--399.

\bibitem{Emam:2007dy_Zprime}
W.~Emam and S.~Khalil, {\it {Higgs and Z-prime phenomenology in B-L extension
  of the standard model at LHC}},  {\em Eur.Phys.J.} {\bf C55} (2007) 625--633,
  [\href{http://xxx.lanl.gov/abs/0704.1395}{{\tt arXiv:0704.1395}}].

\bibitem{Khalil:2010iu_tevBL}
S.~Khalil, {\it {TeV-scale gauged B-L symmetry with inverse seesaw mechanism}},
   {\em Phys.Rev.} {\bf D82} (2010) 077702,
  [\href{http://xxx.lanl.gov/abs/1004.0013}{{\tt arXiv:1004.0013}}].

\bibitem{Iso:2009nw_tevBL}
S.~Iso, N.~Okada, and Y.~Orikasa, {\it {The minimal B-L model naturally
  realized at TeV scale}},  {\em Phys.Rev.} {\bf D80} (2009) 115007,
  [\href{http://xxx.lanl.gov/abs/0909.0128}{{\tt arXiv:0909.0128}}].

\bibitem{Basso:2008iv_phenoBL}
L.~Basso, A.~Belyaev, S.~Moretti, and C.~H. Shepherd-Themistocleous, {\it
  {Phenomenology of the minimal B-L extension of the Standard model: Z' and
  neutrinos}},  {\em Phys.Rev.} {\bf D80} (2009) 055030,
  [\href{http://xxx.lanl.gov/abs/0812.4313}{{\tt arXiv:0812.4313}}].

\bibitem{Okada:2011en}
N.~Okada, M.~U. Rehman, and Q.~Shafi, {\it {Non-Minimal B-L Inflation with
  Observable Gravity Waves}},  {\em Phys.Lett.} {\bf B701} (2011) 520--525,
  [\href{http://xxx.lanl.gov/abs/1102.4747}{{\tt arXiv:1102.4747}}].

\bibitem{Okada:2013vxa}
N.~Okada and Q.~Shafi, {\it {Observable Gravity Waves From U(1)$_{B-L}$ Higgs
  and Coleman-Weinberg Inflation}},
  \href{http://xxx.lanl.gov/abs/1311.0921}{{\tt arXiv:1311.0921}}.

\bibitem{Arcadi:2013qia}
G.~Arcadi, Y.~Mambrini, M.~H.~G. Tytgat, and B.~Zaldivar, {\it {Invisible
  $Z^\prime$ and dark matter: LHC vs LUX constraints}},  {\em JHEP} {\bf 1403}
  (2014) 134, [\href{http://xxx.lanl.gov/abs/1401.0221}{{\tt
  arXiv:1401.0221}}].

\bibitem{Sanchez-Vega:2014rka}
B.~Sánchez-Vega, J.~Montero, and E.~Schmitz, {\it {Complex Scalar DM in a B-L
  Model}},  \href{http://xxx.lanl.gov/abs/1404.5973}{{\tt arXiv:1404.5973}}.

\bibitem{Basak:2013cga}
T.~Basak and T.~Mondal, {\it {Constraining Minimal $U(1)_{B-L}$ model from Dark
  Matter Observations}},  {\em Phys.Rev.} {\bf D89} (2014) 063527,
  [\href{http://xxx.lanl.gov/abs/1308.0023}{{\tt arXiv:1308.0023}}].

\bibitem{Khalil:2008kp}
S.~Khalil and O.~Seto, {\it {Sterile neutrino dark matter in B - L extension of
  the standard model and galactic 511-keV line}},  {\em JCAP} {\bf 0810} (2008)
  024, [\href{http://xxx.lanl.gov/abs/0804.0336}{{\tt arXiv:0804.0336}}].

\bibitem{Okada:2010wd}
N.~Okada and O.~Seto, {\it {Higgs portal dark matter in the minimal gauged
  $U(1)_{B-L}$ model}},  {\em Phys.Rev.} {\bf D82} (2010) 023507,
  [\href{http://xxx.lanl.gov/abs/1002.2525}{{\tt arXiv:1002.2525}}].

\bibitem{Kanemura:2011vm}
S.~Kanemura, O.~Seto, and T.~Shimomura, {\it {Masses of dark matter and
  neutrino from TeV scale spontaneous $U(1)_{B-L}$ breaking}},  {\em Phys.Rev.}
  {\bf D84} (2011) 016004, [\href{http://xxx.lanl.gov/abs/1101.5713}{{\tt
  arXiv:1101.5713}}].

\bibitem{Basso:2012ti}
L.~Basso, O.~Fischer, and J.~van~der Bij, {\it {A natural $Z^\prime$ model with
  inverse seesaw and leptonic dark matter}},  {\em Phys.Rev.} {\bf D87} (2013),
  no.~3 035015, [\href{http://xxx.lanl.gov/abs/1207.3250}{{\tt
  arXiv:1207.3250}}].

\bibitem{Perez:2014qfa}
P.~F. Perez, S.~Ohmer, and H.~H. Patel, {\it {Minimal Theory for Lepto-Baryons
  and Cosmological Implications}},
  \href{http://xxx.lanl.gov/abs/1403.8029}{{\tt arXiv:1403.8029}}.

\bibitem{Hook:2014mla}
A.~Hook, {\it {Inflationary baryogenesis}},
  \href{http://xxx.lanl.gov/abs/1404.0113}{{\tt arXiv:1404.0113}}.

\bibitem{Buchmuller:2013lra}
W.~Buchmüller, V.~Domcke, K.~Kamada, and K.~Schmitz, {\it {The Gravitational
  Wave Spectrum from Cosmological $B-L$ Breaking}},  {\em JCAP} {\bf 1310}
  (2013) 003, [\href{http://xxx.lanl.gov/abs/1305.3392}{{\tt
  arXiv:1305.3392}}].

\bibitem{Kehagias:2014wza}
A.~Kehagias and A.~Riotto, {\it {Remarks about the Tensor Mode Detection by the
  BICEP2 Collaboration and the Super-Planckian Excursions of the Inflaton
  Field}},  \href{http://xxx.lanl.gov/abs/1403.4811}{{\tt arXiv:1403.4811}}.

\bibitem{delAguila:1995rb_2U1}
F.~del Aguila, M.~Masip, and M.~Perez-Victoria, {\it {Physical parameters and
  renormalization of $U(1)_a \times U(1)_b$ models}},  {\em Nucl.Phys.} {\bf
  B456} (1995) 531--549, [\href{http://xxx.lanl.gov/abs/hep-ph/9507455}{{\tt
  hep-ph/9507455}}].

\bibitem{Holdom:1985ag_2U1}
B.~Holdom, {\it {Two U(1)'s and Epsilon Charge Shifts}},  {\em Phys.Lett.} {\bf
  B166} (1986) 196.

\bibitem{delAguila:1988jz_2U1}
F.~del Aguila, G.~Coughlan, and M.~Quiros, {\it {GAUGE COUPLING RENORMALIZATION
  WITH SEVERAL U(1) FACTORS}},  {\em Nucl.Phys.} {\bf B307} (1988) 633.

\bibitem{Appelquist:1974tg}
T.~Appelquist and J.~Carazzone, {\it {Infrared Singularities and Massive
  Fields}},  {\em Phys.Rev.} {\bf D11} (1975) 2856.

\bibitem{EliasMiro:2012ay}
J.~Elias-Miro, J.~R. Espinosa, G.~F. Giudice, H.~M. Lee, and A.~Strumia, {\it
  {Stabilization of the Electroweak Vacuum by a Scalar Threshold Effect}},
  {\em JHEP} {\bf 1206} (2012) 031,
  [\href{http://xxx.lanl.gov/abs/1203.0237}{{\tt arXiv:1203.0237}}].

\bibitem{Lebedev:2012zw}
O.~Lebedev, {\it {On Stability of the Electroweak Vacuum and the Higgs
  Portal}},  {\em Eur.Phys.J.} {\bf C72} (2012) 2058,
  [\href{http://xxx.lanl.gov/abs/1203.0156}{{\tt arXiv:1203.0156}}].

\bibitem{Ade:2013uln}
{\bf Planck Collaboration} Collaboration, P.~Ade et~al., {\it {Planck 2013
  results. XXII. Constraints on inflation}},
  \href{http://xxx.lanl.gov/abs/1303.5082}{{\tt arXiv:1303.5082}}.

\bibitem{NeferSenoguz:2008nn}
V.~N. Senoguz and Q.~Shafi, {\it {Chaotic inflation, radiative corrections and
  precision cosmology}},  {\em Phys.Lett.} {\bf B668} (2008) 6--10,
  [\href{http://xxx.lanl.gov/abs/0806.2798}{{\tt arXiv:0806.2798}}].

\bibitem{Veltman:1980mj}
M.~Veltman, {\it {The Infrared - Ultraviolet Connection}},  {\em Acta
  Phys.Polon.} {\bf B12} (1981) 437.

\bibitem{Einhorn:1992um}
M.~Einhorn and D.~Jones, {\it {The Effective potential and quadratic
  divergences}},  {\em Phys.Rev.} {\bf D46} (1992) 5206--5208.

\bibitem{Khalil:2006yi}
S.~Khalil, {\it {Low scale $B$ - L extension of the Standard Model at the
  LHC}},  {\em J.Phys.} {\bf G35} (2008) 055001,
  [\href{http://xxx.lanl.gov/abs/hep-ph/0611205}{{\tt hep-ph/0611205}}].

\bibitem{Basso:2010jm_U1BL}
L.~Basso, S.~Moretti, and G.~M. Pruna, {\it {A Renormalisation Group Equation
  Study of the Scalar Sector of the Minimal B-L Extension of the Standard
  Model}},  {\em Phys.Rev.} {\bf D82} (2010) 055018,
  [\href{http://xxx.lanl.gov/abs/1004.3039}{{\tt arXiv:1004.3039}}].

\bibitem{Basso:2013vla}
L.~Basso, {\it {Minimal Z' models and the 125 GeV Higgs boson}},  {\em
  Phys.Lett.} {\bf B725} (2013) 322--326,
  [\href{http://xxx.lanl.gov/abs/1303.1084}{{\tt arXiv:1303.1084}}].

\end{thebibliography}\endgroup

%------------------------------------------------------------

\end{document}